\documentclass[aps,prb,a4paper,notitlepage,onecolumn]{revtex4}%
\usepackage{epsfig}
\usepackage{bm}
\usepackage{amsmath}
\usepackage{amsfonts}
\usepackage{amssymb}
\usepackage{graphicx}%
\begin{document}
\title{Spin Pumping and Spin Transfer}
\author{Arne Brataas$^{1}$, Yaroslav Tserkovnyak$^{2}$, Gerrit E. W. Bauer$^{3,4}$,
and Paul J. Kelly$^{5}$}
\affiliation{$^{1}$Department of Physics, Norwegian University of Science and Technology,
N-7491 Trondheim, Norway}
\affiliation{$^{2}$Department of Physics and Astronomy, University of California, Los
Angeles, California, 900095, USA}
\affiliation{$^{3}$Delft University of Technology, Kavli Institute of NanoScience, 2628 CJ
Delft, The Netherlands}
\affiliation{$^{4}$Institute for Materials Research, Tohoku University, Sendai 980-8577, Japan}
\affiliation{$^{5}$Department of Applied Physics, Twente University, Enschede, The Netherlands}

\begin{abstract}
Spin pumping is the emission of a spin current by a magnetization dynamics
while spin transfer stands for the excitation of magnetization by spin
currents. Using Onsager's reciprocity relations we prove that spin pumping and
spin-transfer torques are two fundamentally equivalent dynamic processes in
magnetic structures with itinerant electrons. We review the theory of the
coupled motion of the magnetization order parameter and electron for textured
bulk ferromagnets (\textit{e.g.} containing domain walls) and heterostructures
(such as spin valves). We present first-principles calculations for the
material-dependent damping parameters of magnetic alloys. Theoretical and
experimental results agree in general well.

\end{abstract}
\maketitle
\tableofcontents

\section{Introduction}

\subsection{Technology Pull and Physics Push}

The interaction between electric currents and the magnetic order parameter in
conducting magnetic micro- and nanostructures has developed into a major
subfield in magnetism\cite{Bader}. The main reason is the technological
potential of magnetic devices based on transition metals and their alloys that
operate at ambient temperatures. Examples are current-induced tunable
microwave generators (spin-torque oscillators)\cite{Silva,Braganca}, and
non-volatile magnetic electronic architectures that can be randomly read,
written or programmed by current pulses in a scalable manner\cite{Matsunaga}.
The interaction between currents and magnetization can also cause undesirable
effects such as enhanced magnetic noise in read heads made from magnetic
multilayers\cite{Nagasaka}. While most research has been carried out on
metallic structures, current-induced magnetization dynamics in
semiconductors\cite{Awschalom} or even insulators\cite{Kajiwara} has been
pursued as well.

Physicists have been attracted in large numbers to these issues because on top
of the practical aspects the underlying phenomena are so fascinating.
Berger\cite{Berger:prb96} and Slonczewski\cite{Slonczewski:jmmm96} are in
general acknowledged to have started the whole field by introducing the
concept of current-induced magnetization dynamics by the transfer of spin. The importance of their work was fully appreciated only after
experimental confirmation of the predictions in multi-layered
structures\cite{Tsoi:prl98,Myers:sci99}. The reciprocal effect, \textit{i.e.}.
the generation of currents by magnetization dynamics now called\textit{ spin
pumping}, has been expected long ago\cite{Janossy:prl76,Silsbee:prb79}, but
it took some time before Tserkovnyak \textit{et al}.
\cite{Tserkovnyak:prl02,Tserkovnyak:rmp05} developed a rigorous theory of
spin-pumping for magnetic multi-layers, including the associated increased
magnetization damping\cite{Mizukami,Urban:prl01,Heinrich:prl03}.

\subsection{Discrete versus Homogeneous}

Spin-transfer torque and spin pumping in magnetic metallic multi-layers are by
now relatively well understood and the topic has been covered by a number of review
articles \cite{Tserkovnyak:rmp05,Brataas:pr06,Ralph:jmmm08}. It can be
understood very well in terms of a time-dependent extension of
magneto-electronic circuit theory\cite{Brataas:prl00,Brataas:pr06}, which
corresponds to the assumption of spin diffusion in the bulk and quantum
mechanical boundary conditions at interfaces. Random matrix
theory\cite{Waintal:prb00} can be shown to be equivalent to circuit
theory\cite{Bauer:prb03,Brataas:pr06,Rychkov:prl09}. The technologically
important current-induced switching in magnetic tunnel junctions has recently
been the focus of attention\cite{Sun:jmmm08}. Tunnel junctions limit the
transport such that circuit issues are less important, whereas the
quantum-mechanical nature of the tunneling process becomes essential. We will
not review this issue in more detail here.

The interaction of currents and magnetization in continuous magnetization
textures has also attracted much interest, partly due to possible applications
such as nonvolatile shift registers\cite{Parkin:sc08}. From a formal point of
view the physics of current-magnetization interaction in a continuum poses new
challenges as compared to heterostructures with atomically sharp interfaces.
In magnetic textures such as magnetic domain walls, currents interact over
length scales corresponding to the wall widths that are usually much longer
than even the transport mean-free path. Issues of the in-plane \textit{vs.}
magnetic-field like torque\cite{Zhang:prl04} and the spin-motive force in
moving magnetization textures\cite{Barnes:prl07} took some time to get sorted
out, but the understanding of the complications associated with continuous
textures has matured by now. There is now general consensus about the physics
of current-induced magnetization excitations and magnetization dynamics induced
currents\cite{Tatara:pr08,Beach:jmmm08}. Nevertheless, the similarities and
differences of spin torque and spin pumping in discrete and continuous
magnetic systems has to our knowledge never been discussed in a coherent
fashion. It has also only recently been realized that both phenomena are directly related, since they reflect identical microscopic correlations
according to the Onsager reciprocity
relations\cite{Tserkovnyak:prb08,Hals:prl09,Bauer:prb10}.

\subsection{This Chapter}

In this Chapter, we (i) review the basic understandings of spin transfer
torque \textit{vs.} spin pumping and (ii) knit together our understanding of
both concepts for heterogeneous and homogeneous systems. We discuss the
general phenomenology guided by Onsager's reciprocity in the linear response
regime\cite{deGroot:book52}. We will compare the in- and out-of-plane spin
transfer torques at interfaces as governed by the real and imaginary part of
the so-called spin-mixing conductances with that in textures, which are
usually associated with the adiabatic torque and its dissipative
correction\cite{Zhang:prl04}, usually described by a dimensionless factor
$\beta$ in order to stress the relation with the Gilbert damping constant
$\alpha$. We argue that the spin pumping phenomenon at interfaces between
magnets and conductors is identical to the spin-motive force due to
magnetization texture dynamics such as moving domain walls\cite{Barnes:prl07}.
We emphasize that spin pumping is on a microscopic level identical to the spin
transfer torque, thus arriving at a significantly simplified conceptual
picture of the coupling between currents and magnetization. We also point out
that we are not limited to a phenomenological description relying on fitting
parameters by demonstrating that the material dependence of crucial parameters
such as $\alpha$ and $\beta$\textit{ }can be computed from first principles. 

\section{Phenomenology}

In this Section we explain the basics physics of spin-pumping and
spin-transfer torques, introduce the dependence on material and externally
applied parameters, and prove their equivalence in terms of Onsager's
reciprocity theorem. 

\subsection{Mechanics}

On a microscopic level electrons behave as wave-like Fermions with quantized
intrinsic angular momentum. However, in order to understand the electron wave
packets at the Fermi energy in high-density metals and the collective motion
of a large number of spins at not too low temperatures classical analogues can
be useful.

Spin transfer torque and spin pumping are on a fundamental level mechanical
phenomena that can be compared with the game of billiards, which is all about
the transfer of linear and angular momenta between the balls and cushions. A
skilled player can use the cue to transfer velocity and spin to the billiard
ball in a controlled way. The path of the spinning ball is governed by the
interaction with the reservoirs of linear and angular momentum (the cushions
and the felt/baize) and with other balls during collisions. A ball that for
instance hits the cushion at normal angle with top or bottom spin will reverse
its rotation and translation, thereby transferring twice its linear and
angular moment to the frame of the billiard.

Since the work by Barnett\cite{Barnett} and Einstein-de
Haas\cite{EinsteindeHaas} almost a century ago, we know that magnetism is
caused by the magnetic moment of the electron, which is intimately related
with its mechanical angular momentum. How angular momentum transfer occurs
between electrons in magnetic structures can be imagined mechanically: just
replace the billiard balls by spin polarized electrons and the cushion by a
ferromagnet. Good metallic interfaces correspond to a cushion with high
friction. The billiard ball reverses angular and linear momentum, whereas the
electron is reflected with a spin flip. While the cushion and the billiard
table absorb the angular momentum, the magnetization absorbs the spin angular
momentum. The absorbed spins correspond to a torque that, if exceeding a
critical value, will set the magnetization into motion. Analogously, a time-dependent
magnetization injects net angular momentum into a normal metal contact. This
\textquotedblleft spin pumping\textquotedblright\ effect, \textit{i.e.} the
main topic of this chapter, can be also visualized mechanically: a billiard
ball without spin will pick up\ angular momentum under reflection if the
cushion is rotating along its axis.

\subsection{Spin-transfer Torque and Spin-pumping}

Ferromagnets do not easily change the modulus of the magnetization vector due
to large exchange energy costs. The low-energy excitations, so-called spin
waves or magnons, only modulate the magnetization direction with respect to
the equilibrium magnetization configuration. In this regime the magnetization
dynamics of ferromagnets can be described by the Landau-Lifshitz-Gilbert (LLG)
equation,%
\begin{equation}
\mathbf{\dot{m}}=-\gamma\mathbf{m}\times\mathbf{H}_{\mathrm{eff}}%
+\tilde{\alpha}\mathbf{m}\times\mathbf{\dot{m}}, \label{LLG}%
\end{equation}
where $\mathbf{m}\left(  \mathbf{r},t\right)  $ is a unit vector along the
magnetization direction, $\mathbf{\dot{m}}=\partial\mathbf{m}/\partial t$,
$\gamma=g^{\ast}\mu_{B}/\hbar>0$ is (minus) the gyro-magnetic ratio in terms
of the effective $g$-factor and the Bohr magneton $\mu_{B}$, and
$\tilde{\alpha}$ is the Gilbert damping tensor that determines the
magnetization dissipation rate. Under isothermal conditions the effective
magnetic field $\mathbf{H}_{\mathrm{eff}}=-\delta F\left[  \mathbf{m}\right]
/\delta(M_{s}\mathbf{m})$ is governed by the magnetic\ free energy $F$ and
$M_{s}$ is the saturation magnetization. We will consider both spatially
homogeneous and inhomogeneous situations. In the former case, the
magnetization is constant in space (macrospin), while the torques are applied
at the interfaces. In the latter case, the effective magnetic field
$\mathbf{H}_{\mathrm{eff}}$ also includes a second order spatial gradient
arising from the (exchange) rigidity of the magnetization and torques as well
as motive forces that are distributed in the ferromagnet.

Eq. (\ref{LLG}) can be rewritten in the form of the Landau-Lifshitz (LL)
equation:
\begin{equation}
\left(  1+\tilde{\alpha}^{2}\right)  \mathbf{\dot{m}}=-\gamma\mathbf{m}%
\times\mathbf{H}_{\mathrm{eff}}-\gamma\tilde{\alpha}\mathbf{m}\times\left(
\mathbf{m}\times\mathbf{H}_{\mathrm{eff}}\right)  . \label{LL}%
\end{equation}
Additional torques due to the coupling between currents and magnetization
dynamics should be added to the right-hand side of the LLG or LL equation, but
some care should be exercised in order to keep track of dissipation in a
consistent manner. In our approach the spin-pumping and spin-transfer torque
contributions are most naturally added to the LLG equation (\ref{LLG}), but we
will also make contact with the LL equation\ (\ref{LL}) while exploring the
Onsager reciprocity relations.

In the remaining part of this section we describe the extensions of the LLG equation
due to spin-transfer and spin-pumping torques for discrete and bulk
systems in Sec. \ref{Discrete} and Sec. \ref{Continuous}, respectively. In the
next section we demonstrate in more detail how spin-pumping and spin-transfer
torque are related by Onsager reciprocity relations for both discrete and
continuous systems.

\subsubsection{Discrete Systems\label{Discrete}}

Berger and Slonczewski predicted that in spin-valve structures with current
perpendicular to the interface planes (CPP) a dc current can excite and even
reverse the reverse the relative magnetization of magnetic layers separated by
a normal metal spacer\cite{Berger:prb96,Slonczewski:jmmm96}. The existence of
this phenomenon has been amply confirmed by
experiments\cite{Tsoi:prl98,Myers:sci99,Grollier:apl01,Kiselev:nat03,Ozyilmaz:prl03,Krivorotov:scie05,Ralph:jmmm08,Cui:10}%
. We can understand current-induced magnetization dynamics from first
principles in terms of the coupling of spin-dependent transport with the
magnetization. In a ferromagnetic metal majority and minority electron spins
have often very different electronic structures. Spins that are polarized
non-collinear with respect to the magnetization direction are not eigenstates
of the ferromagnet, but can be described as a coherent linear combination of
majority and minority electron spins at the given energy shell. If injected at
an interface, these states precess on time and length scales that depend on
the orbital part of the wave function. In high electron-density transition
metal ferromagnets like Co, Ni, and Fe a large number of wave vectors are
available at the Fermi energy. A transverse spin current injected from a
diffuse reservoir generates a large number of wave functions oscillating with
different wave length that lead to efficient destructive interference or
decoherence of the spin momentum. Beyond a transverse magnetic coherence
length, which in these materials is of the order of the Fermi wave length,
typically around 1 nm, a transversely polarized spin current cannot
persist.\cite{Brataas:prl00} This destruction of transverse angular momentum
is per definition equal to a torque. Slonczewski's spin-transfer torque is
therefore equivalent to the absorption of a spin current at an interface
between a normal metal and a ferromagnet whose magnetization is transverse to
the spin current polarization. Each electron carries an electric charge $-e$
and an angular momentum of $\pm\hbar/2$. The loss of transverse spin angular
momentum at the normal metal-ferromagnet interface is therefore $\hbar\left[
\mathbf{I}_{s}-\left(  \mathbf{I}_{s}\cdot\mathbf{m}\right)  \mathbf{m}%
\right]  /(2e),$ where the spin-current $\mathbf{I}_{s}$ is measured in the
units of an electrical current, \emph{e.g.} in Ampere. In the macrospin
approximation the torque has to be shared with all magnetic moments or
$M_{s}\mathcal{V}$ of the ferromagnetic particle or film with volume
$\mathcal{V}$. The torque on magnetization equals the rate of change of the
total magnetic moment of the magnet $\partial\left(  \mathbf{m}M_{s}%
\mathcal{V}\right)  _{\mathrm{stt}}/\partial t,$ which equals the spin current
absorption\cite{Slonczewski:jmmm96} .The rate of change of the magnetization
direction therefore reads:
\begin{equation}
\boldsymbol{\tau}_{\mathrm{stt}}=\left(  \frac{\partial\mathbf{m}}{\partial
t}\right)  _{\mathrm{stt}}=-\frac{\gamma\hbar}{2eM_{s}\mathcal{V}}%
\mathbf{m}\times\left(  \mathbf{m\times I}_{s}\right)
.\label{Slonczewski_torqu}%
\end{equation}

\begin{figure}[htbp]
\includegraphics[width=0.5\columnwidth]{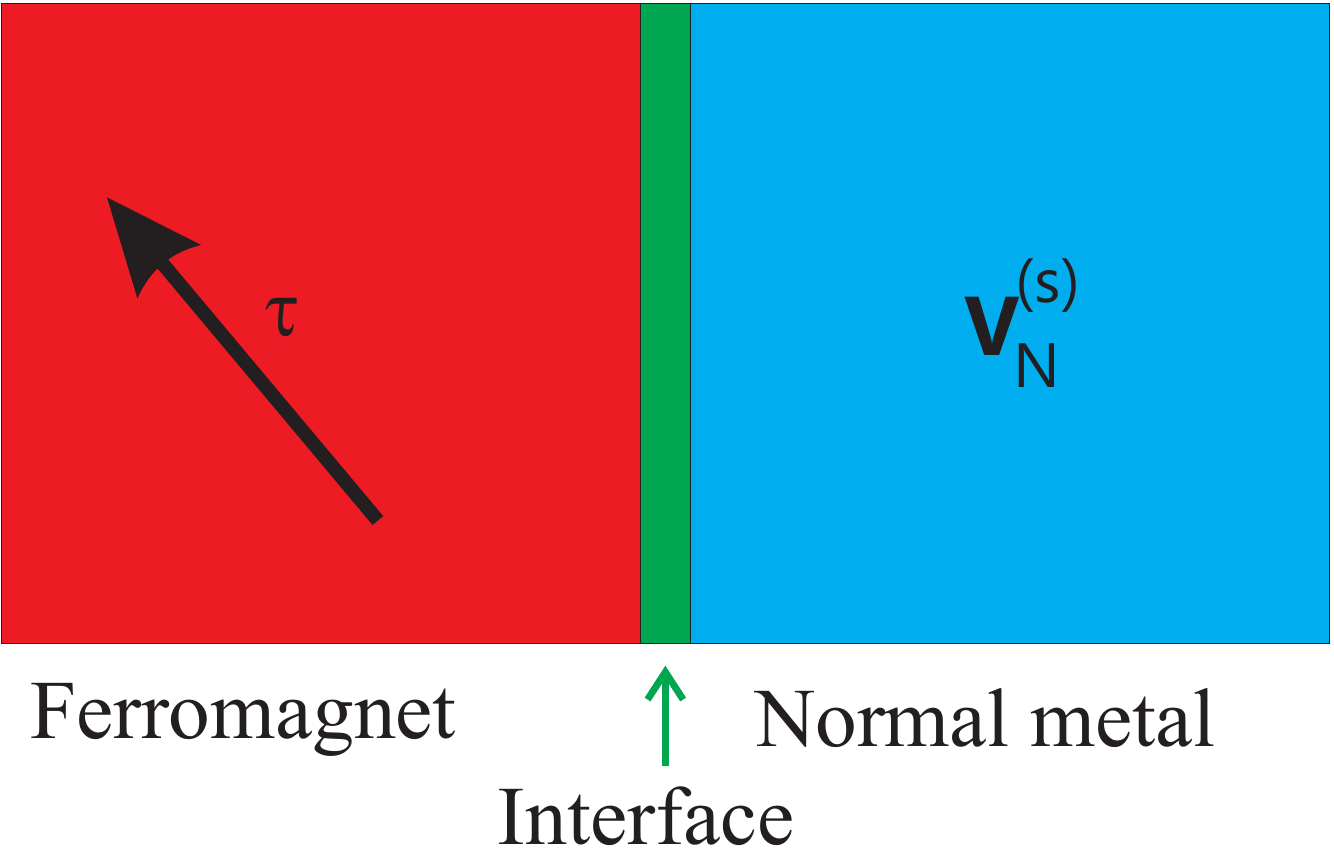} \caption{Illustration of
the spin-transfer torque in layered normal metal$|$ferromagnet system. A spin
accumulation $\mathbf{V}_{N}^{(s)}$ in the normal metal induces a
spin-transfer torque ${\boldsymbol{\tau}}_{\mathrm{stt}}$ on the ferromagnet.}%
\label{fig:spintorque}%
\end{figure}

We still need to evaluate the spin current that can be generated,
\textit{e.g}., by the inverse spin Hall effect in the normal metal or optical
methods. Here we concentrate on the layered normal metal-ferromagnet systems
in which the current generated by an applied bias is polarized by a second
highly coercive magnetic layer as in the schematic Fig. \ref{fig:spintorque}.
Magnetoelectronic circuit theory is especially suited to handle such a problem
\cite{Brataas:prl00} For simplicity we disregard here extrinsic dissipation of spin
angular momentum due to spin-orbit coupling and disorder, which can taken into
account when the need arises\cite{Kovalev:prb02,Hals:epl10}. We allow for a
non-equilibrium magnetization or spin accumulation $\mathbf{V}_{N}^{(s)}$ in
the normal metal layer. $\mathbf{V}_{N}^{(s)}$ is a vector pointing in the
direction of the local net magnetization, whose modulus $V_{N}^{(s)}$ is the
difference between the differences in electric potentials (or electrochemical
potentials divided by $2e$) of both spin species. Including the charge
accumulation $V_{N}^{(c)}$ (local voltage), the potential experienced by a
spin-up (spin-down) electron along the direction of the spin accumulation in
the normal metal is $V_{N}^{\uparrow}=V_{N}^{(c)}+V_{N}^{(s)}$ $\left(
V_{N}^{\downarrow}=V_{N}^{(c)}-V_{N}^{(s)}\right)  $. Inside a ferromagnet,
the spin accumulation must be aligned to the magnetization direction
$\mathbf{V}_{F}^{(s)}=\mathbf{m}V_{F}^{(s)}$. Since $V_{F}^{(s)}$ does not
directly affect the spin-transfer torque at the interface we disregard it for
convenience here (see Ref.
\onlinecite{Brataas:pr06}
for a complete treatment), but retain the charge accumulation $V_{F}^{(c)}$.
We can now compute the torque at the interface between a normal metal and a
ferromagnet arising from a given spin accumulation $\mathbf{V}_{N}^{(s)}$.
Ohm's Law for the spin-current projections aligned ($I_{\uparrow}$) and
anti-aligned ($I_{\downarrow}$)\ to the magnetization direction then
read\cite{Brataas:prl00,Brataas:prb06} (positive currents correspond to charge
flowing from the normal metal towards the ferromagnet)
\begin{align}
I_{\uparrow} &  =G_{\uparrow}\left[  \left(  V_{N}^{(c)}-V_{F}^{(c)}\right)
+\mathbf{m\cdot}\left(  \mathbf{V}_{N}^{(s)}-\mathbf{m}V_{F}^{(c)}\right)
\right]  ,\label{Iu}\\
I_{\downarrow} &  =G_{\downarrow}\left[  \left(  V_{N}^{(c)}-V_{F}%
^{(c)}\right)  -\mathbf{m\cdot}\left(  \mathbf{V}_{N}^{(s)}-\mathbf{m}%
V_{F}^{(c)}\right)  \right]  .\label{Id}%
\end{align}
where $G_{\uparrow}$ and $G_{\downarrow}$ are the spin-dependent interface
conductances. The total charge current $I^{(c)}=I_{\uparrow}+I_{\downarrow}$,
is continuous across the interface, $I_{N}^{(c)}=I_{F}^{(c)}=I^{(c)}$. The
(longitudinal) spin current defined by Eqs. (\ref{Iu}) and (\ref{Id}) $\left(
I_{\uparrow}-I_{\downarrow}\right)  \mathbf{m}$ is polarized along the
magnetization direction. The transverse part of the spin current can be
written as the sum of two vector components in the space spanned by the
$\mathbf{m}$,$\mathbf{V}_{N}^{(s)}$ plane as well as its normal. The total
spin current on the normal metal side close to the interface
reads\cite{Brataas:prl00,Brataas:pr06}:
\begin{equation}
\mathbf{I}_{N}^{(s,\mathrm{bias})}=\left(  I_{\uparrow}-I_{\downarrow}\right)
\mathbf{m}-2G_{\perp}^{(R)}\mathbf{m}\times\left(  \mathbf{m}\times
\mathbf{V}_{N}^{(s)}\right)  -2G_{\perp}^{(I)}\left(  \mathbf{m}%
\times\mathbf{V}_{N}^{(s)}\right)  ,\label{spincurrent}%
\end{equation}
where $G_{\perp}^{(R)}$ and $G_{\perp}^{(I)}$ are two independent transverse
interface conductances. $\mathbf{I}_{N}^{(s,\mathrm{bias})}$ is driven by the
external bias $\mathbf{V}_{N}^{(s)}$ and should be distinguished from the pumped spin
current addressed below. $\left(  R\right)  $ and $\left(  I\right)  $ refer
to the real and imaginary parts of microscopic expression for these
\textquotedblleft spin mixing\textquotedblright\ interface conductances
$G_{\uparrow\downarrow}=G_{\perp}^{(R)}+iG_{\perp}^{(I)}$.

The transverse components are absorbed in the ferromagnet within a very thin
layer. Detailed calculations show that transverse spin-current absorption in
the ferromagnet happens within a nanometer from the interface, where disorder
suppresses any residual oscillations that survived the above-mentioned
destructive interference in ballistic structures\cite{Zwierzycki:prb05}.
Spin-transfer in transition metal based multilayers is therefore an interface
effect, except in ultrathin ferromagnetic films\cite{Kovalev:prb06}. As discussed
above, the divergence of the transverse spin current at the interface gives
rise to the torque%
\begin{equation}
{\boldsymbol{\tau}}_{\mathrm{stt}}^{(\mathrm{bias})}=-\frac{\gamma\hbar
}{eM_{s}\mathcal{V}}\left[  G_{\perp}^{(R)}\mathbf{m}\times\left(
\mathbf{m}\times\mathbf{V}_{N}^{(s)}\right)  +G_{\perp}^{(I)}\left(
\mathbf{m}\times\mathbf{V}_{N}^{(s)}\right)  \right]  .
\label{spintransfertorque}%
\end{equation}
Adding this torque to the Landau-Lifshitz-Gilbert equation leads to the
Landau-Lifshitz-Gilbert-Slonczewski (LLGS) equation
\begin{equation}
\mathbf{\dot{m}}=-\gamma\mathbf{m}\times\mathbf{H}_{\mathrm{eff}%
}+{\boldsymbol{\tau}}_{\mathrm{stt}}^{(\mathrm{bias})}+\alpha\mathbf{m}%
\times\mathbf{\dot{m}.} \label{LLGS}%
\end{equation}
The first term in Eq.\ (\ref{spintransfertorque}) is the (Slonczewski) torque
in the $\left(  \mathbf{m},\mathbf{V}_{N}^{(s)}\right)  $ plane, which
resembles the Landau-Lifshitz damping in Eq.\ (\ref{LL}). When the
spin-accumulation $\mathbf{V}_{N}^{(s)}$ is aligned with the effective
magnetic field $\mathbf{H}_{\mathrm{eff}}$, the Slonczewski torque effectively
enhances the damping of the ferromagnet and stabilizes the magnetization
motion towards the equilibrium direction. On the other hand, when
$\mathbf{V}_{N}^{(s)}$ is antiparallel to $\mathbf{H}_{\mathrm{eff}}$, this
torque opposes the damping. When exceeding a critical value it leads to
precession or reversal of the magnetization. The second term in
Eq.\ (\ref{spintransfertorque}) proportional to $G_{\perp}^{(I)}$ modifies the
magnetic field torque and precession frequency. While the in-plane torque
leads to dissipation of the spin accumulation, the out-of-plane torque induces
a precession of the spin accumulation in the ferromagnetic exchange field
along $\mathbf{m}$. It is possible to implement the spin-transfer torque into
the Landau-Lifshitz equation, but the conductance parameters differ from those
in Eq.\ (\ref{spintransfertorque}).

\begin{figure}[ptb]
\includegraphics[width=0.5\columnwidth]{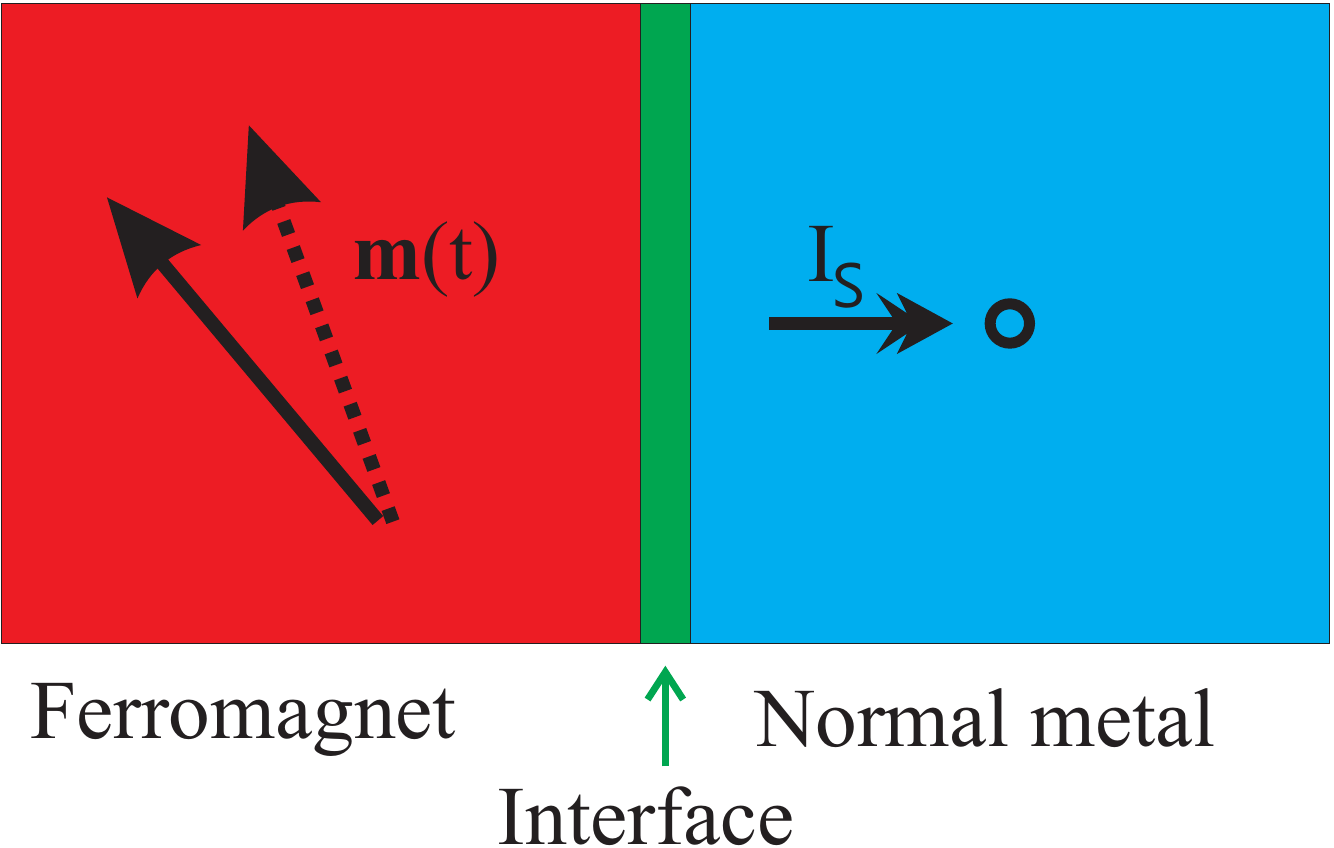} \caption{Spin-pumping in
normal metal$|$ferromagnet systems. A dynamical magnetization
\textquotedblleft pumps\textquotedblright\ a spin current $\mathbf{I}^{(s)}$
into an adjacent normal metal.}%
\label{fig:spinpumping}%
\end{figure}

Since spin currents can move magnetizations, it is natural to consider the
reciprocal effect, \textit{viz}. the generation of spin currents by
magnetization motion. It was recognized in the 1970's that spin dynamics is
associated with spin currents in normal metals. Barnes\cite{Barnes:jpf74}
studied the dynamics of localized magnetic moments embedded in a conducting
medium. He showed that the dynamic susceptibility in diffuse media is limited
by the spin-diffusion length. Janossy and Monod\cite{Janossy:prl76} and
Silsbee \textit{et al}.\cite{Silsbee:prb79} postulated a coupling between a
dynamic ferromagnetic magnetization and a spin accumulation in adjacent normal
metals in order to explain that microwave transmission through normal metal
foils is enhanced by a coating with a ferromagnetic layer. The scattering
theory for spin currents induced by magnetization dynamics was developed by
Tserkovnyak \textit{et al}.\cite{Tserkovnyak:prl02} on the basis of the theory
of adiabatic quantum pumping\cite{Buttiker}, hence the name \textquotedblleft
spin pumping\textquotedblright. Theoretical results were confirmed by the
agreement of the spin-pumping induced increase of the Gilbert damping with
experiments by Mizukami \textit{et al} and Heinrich et al.\cite{Mizukami,Urban:prl01,Heinrich:prl03}. A schematic picture of spin-pumping in normal$|$ferromagnet systems is shown in Fig.\ \ref{fig:spinpumping}.
At not too high excitations and temperatures, the ferromagnetic
dynamics conserves the modulus of the magnetization $M_{s}\mathbf{m}$.
Conservation of angular momentum then implies that the spin current
$\mathbf{I}_{N}^{(s,\mathrm{pump})}$ pumped out of the ferromagnet has to be
polarized perpendicularly to $\mathbf{m,}$ \textit{viz.} $\mathbf{m\cdot
I}_{N}^{(s,\mathrm{pump})}=0$. Furthermore, the adiabatically pumped spin
current is proportional to $\left\vert \mathbf{\dot{m}}\right\vert $. Under
these conditions, therefore,\cite{Tserkovnyak:prl02,Tserkovnyak:rmp05}
\begin{equation}
\frac{e}{\hbar}\mathbf{I}_{N}^{(s,\mathrm{pump})}=G_{\bot}^{\prime(R)}\left(
\mathbf{m}\times\mathbf{\dot{m}}\right)  +G_{\bot}^{\prime(I)}\mathbf{\dot{m}%
},\label{spin-pumping}%
\end{equation}
where $G_{\bot}^{^{\prime}R}$ and $G_{\bot}^{^{\prime}I}$ are two transverse
conductances that depend on the materials. Here the sign is defined to be
negative when $\mathbf{I}_{N}^{(s,\mathrm{pump})}$ implies loss of angular
momentum for the ferromagnet. For $\left\vert \mathbf{\dot{m}}\right\vert
\neq0,$ the right-hand side of the LLGS equation (\ref{LLGS})\ must be
augmented by Eq.\ (\ref{spin-pumping}). The leakage of angular momentum leads
\textit{e.g.}\ to an enhanced Gilbert damping\cite{Mizukami,Urban:prl01,Heinrich:prl03}.

Onsager's reciprocity relations dictate that conductance parameters in
thermodynamically reciprocal processes must be identical when properly
normalized. We prove below that spin-transfer torque (\ref{spintransfertorque}%
) and spin pumping (\ref{spin-pumping})\ indeed belong to this category and
must be identical, \textit{viz.} $G_{\bot}^{(R)}=G_{\bot}^{\prime(R)}$ and
$G_{\bot}^{(I)}=G_{\bot}^{\prime(I)}$. Spin-transfer torque and spin-pumping
are therefore opposite sides of the same coin, at least in the linear response
regime. Since spin-mixing conductance parameters governing both processes are
identical, an accurate measurement of one phenomenon is sufficient to quantify
the reciprocal process. Magnetization dynamics induced by the spin-transfer
torque are not limited to macrospin excitations and experiments are carried
out at high current levels that imply heating and other complications. On the
other hand, spin-pumping can be directly detected by the line-width broadening
of FMR spectra of thin multilayers. In the absence of two-magnon scattering
phenomena and a sufficiently strong static magnetic field, FMR excites only
the homogeneous macrospin mode, allowing the measurement of the transverse
conductances $G_{\bot}^{\prime(R)}$ and, in principle, $G_{\bot}^{\prime(I)}$. $G_{\bot
}^{\prime(I)}.$ Experimental results and first-principles
calculations\cite{Tserkovnyak:prl02,Tserkovnyak:rmp05} agree quantitatively
well. Rather than attempting to measure these parameters by current-induced
excitation measurements, the values $G_{\bot}^{\prime(R)}$ and $G_{\bot
}^{\prime(I)}$ should be inserted, concentrating on other parameters when
analyzing these more complex magnetization phenomena. Finally we note that
spin mixing conductance parameters can be derived as well from static
magnetoresistance measurements in spin valves\cite{Kovalev:prb06} or by
detecting the spin current directly by the inverse spin Hall
effect\cite{Saitoh:apl06,Czechka:10}.

\subsubsection{Continuous Systems\label{Continuous}}

The coupling effects between (spin-polarized) electrical currents and
magnetization dynamics also exist in magnetization textures of bulk metallic
ferromagnets. Consider a magnetization that adiabatically varies its direction
in space. The dominant contribution to the spin-transfer torque can be
identified as a consequence of violation of angular momentum conservation: In
a metallic ferromagnet, a charge current is spin polarized along the
magnetization direction to leading order in the texture gradients. In the
bulk, \textit{i.e.} separated from contacts by more than the spin-diffusion
length, the current polarization is $P=(\sigma_{\uparrow}-\sigma_{\downarrow
})/(\sigma_{\uparrow}+\sigma_{\downarrow}),$ in terms of the ratio of the
conductivities for majority and minority electrons, where we continue to
measure spin currents in units of electric currents. We first disregard
spin-flip processes that dissipate spin currents to the lattice. To zeroth
order in the gradients, the spin current $\mathbf{j}^{(s)}$ flowing is a
specified (say $x$-) direction at position $\mathbf{r}$ is polarized along the
local magnetization, $\mathbf{j}^{(s)}\left(  \mathbf{r}\right)
=\mathbf{m}(\mathbf{r})j^{(s)}(\mathbf{r})$. The gradual change of the
magnetization direction corresponds to a divergence of the angular momentum of
the itinerant electron subsystem, $\partial_{x}\mathbf{j}^{(s)}=j^{(s)}%
\partial_{x}\mathbf{m}+\mathbf{m}\partial_{x}j^{(s)}$, where the latter term
is aligned with the magnetization direction and does not contribute to the
magnetization torque. This change of spin current does not
leave the electron system but flows into the magnetic order, thus inducing a
torque on the magnetization. This process does not cause any dissipation and
the torque is reactive, as can be seen as well from its time reversal
symmetry. To first order in the texture gradient, or adiabatic limit, and for
arbitrary current directions\cite{Volovik:jpc87,Tatara:prl04}
\begin{equation}
{\boldsymbol{\tau}}_{\mathrm{stt}}^{(\mathrm{bias})}(\mathbf{r})=\frac
{g^{\ast}\mu_{B}P}{2eM_{s}}\left(  \mathbf{j}\cdot\nabla\right)
\mathbf{m}\,,\label{torque_adiabatic_inplane}%
\end{equation}
where $\mathbf{j}$ is the charge current density vector and the superscript
\textquotedblleft bias\textquotedblright\ indicates that the torque is induced
by a voltage bias or electric field. From symmetry arguments another torque
should exist that is normal to Eq. (\ref{torque_adiabatic_inplane}), but still
perpendicular to the magnetization and proportional to the lowest order in its
gradient. Such a torque is dissipative, since it changes sign under time
reversal. For isotropic systems, we can parameterize the out-of-plane torque
by a dimensionless parameter $\beta$ such that the total torque reads
\cite{Zhang:prl04,Thiaville:epl05},
\begin{equation}
{\boldsymbol{\tau}}_{\mathrm{stt}}^{(\mathrm{bias})}(\mathbf{r})=\frac
{g^{\ast}\mu_{B}}{2eM_{s}}\sigma P\left[  \left(  \mathbf{E\cdot\nabla
}\right)  \mathbf{m}+\beta\mathbf{m\times}\left(  \mathbf{E\cdot\nabla
}\right)  \mathbf{m}\right]  ,\label{t_stt}%
\end{equation}
we have used Ohm's law, $\mathbf{j}=\sigma\mathbf{E}$. In the adiabatic limit,
\textit{i.e.} to the first order in the gradient of the magnetization
$\partial_{i}m_{j}$, the spin-transfer torque Eq.\ (\ref{t_stt}) describes how
the magnetization dynamics is affected by currents in isotropic ferromagnets.

Analogous to discrete systems, we may expect a process reciprocal to
(\ref{t_stt}) in ferromagnetic textures similar to the spin pumping at
interfaces. Since we are now operating in a ferromagnet, a pumped spin current
is transformed into a charge current. To leading order a time-dependent
texture is expected to pump a current proportional to the rate of change of
the magnetization direction and the gradient of the magnetization texture. For
isotropic systems, we can express the expected charge current as
\begin{equation}
j_{i}^{(\text{\textrm{pump}})}=\frac{\hbar}{2e}\sigma P^{\prime}\left[
\mathbf{m\times\partial}_{i}\mathbf{m+}\beta^{\prime}\mathbf{\partial}%
_{i}\mathbf{m}\right]  \cdot\mathbf{\dot{m},}\label{currentpumpingcont}%
\end{equation}
where $P^{\prime}$ is a polarization factor and $\beta^{\prime}$ an
out-of-plane contribution. Note that we have here been assuming a strong
spin-flip rate so that the spin-diffusion length is much smaller than the
typical length of the magnetization texture. Volovik considered the opposite limit of weak spin-dissipation and kept
track of currents in two independent spin bands\cite{Volovik:jpc87}. In that
regime he derived the first term in (\ref{currentpumpingcont}), proportional to
$P^{\prime}$ and proved that $P=P^{\prime}$. This results was re-derived by
Barnes and Maekawa\cite{Barnes:prl07}. The last term, proportional to the
$\beta$-factor was first discussed by Duine in
Ref.\ (\onlinecite{Duine:prb08}) for a mean-field model, demonstrating that
$\beta=\beta^{\prime}$. More general textures and spin relaxation regimes were
treated by Tserkovnyak and Mecklenburg\cite{Tserkovnyak:prb08}. In the
following we demonstrate by the Onsager reciprocity relations that the
coefficients appearing in the spin-transfer torques (\ref{t_stt}) are
identical to those in the pumped current (\ref{currentpumpingcont}),
\textit{i.e}. $P=P^{\prime}$ and $\beta=\beta^{\prime}$.

The proposed relations for the spin-transfer torques and pumped current in
continuous systems form a local relationship between torques, current, and
electric and magnetic fields. For ballistic systems, this is not satisfied
since the current at one spatial point depends on the electric field in the
whole sample or global voltage bias and not just on the local electric field. The local assumption also breaks down in other
circumstances. The long-range magnetic dipole interaction typically breaks a
ferromagnet into uniform domains. The magnetization gradually changes in the
region between the domains, the domain wall. When the domain wall width is
smaller than the phase coherence length or the mean free path, one should
replace the local approach by a global strategy for magnetization textures in
which the dynamics is characterized by one or more dynamic (soft)\ collective
coordinates $\left\{  \xi_{a}(\tau)\right\}  $ that are allowed to vary
(slowly) in time
\begin{equation}
\mathbf{m}(\mathbf{r}\tau)=\mathbf{m}_{\mathrm{st}}(\mathbf{r};\left\{
\xi_{a}(\tau)\right\}  ),\label{Collective_coordinate}%
\end{equation}
where $\mathbf{m}_{\mathrm{st}}$ is a static description of the texture. In
order to keep the discussion simple and transparent we disregard
thermoelectric effects, which can be important in principle\cite{Bauer:ssc10}.
The thermodynamic forces are $-\partial F/\partial\xi_{a}$, where $F$ is the free energy as well as the bias
voltage across the sample $V$. In linear response the rate of change of the
dynamic collective coordinates and the charge current in the system are
related to the thermodynamic forces $-\partial F\mathbf{/\partial\xi}$ and $V$
by a response matrix
\begin{equation}
\left(
\begin{array}
[c]{c}%
\mathbf{\dot{\xi}}\\
I
\end{array}
\right)  =\left(
\begin{array}
[c]{cc}%
\tilde{L}_{\xi\xi} & \tilde{L}_{\xi I}\\
\tilde{L}_{I\xi} & \tilde{L}_{II}%
\end{array}
\right)  \left(
\begin{array}
[c]{c}%
-\partial F\mathbf{/\partial\xi}\\
V
\end{array}
\right)  ,
\end{equation}
where $\tilde{L}_{\xi V}$ describes the bias voltage-induced torque and
$\tilde{L}_{I\xi}$ the current pumped by the moving magnetization texture.
These expressions are general and includes \textit{e.g.} effects of spin-orbit
interaction. Onsager's reciprocity relations imply $\tilde{L}_{I\xi_{i}%
}\{\mathbf{m},\mathbf{H}\}=\tilde{L}_{\xi_{i}I}\{-\mathbf{m},-\mathbf{H}\}$ or
$\tilde{L}_{I\xi_{i}}\{\mathbf{m},\mathbf{H}\}=\tilde{L}_{\xi_{i}%
I}\{-\mathbf{m},-\mathbf{H}\}$ depending on how the collective coordinates
transform under time-reversal. The coefficient $\tilde{L}_{I\xi}$ can be
easily expressed in terms of the scattering theory of adiabatic pumping as
discussed below. This strategy was employed to demonstrate for (Ga,Mn)As that
the spin-orbit interaction can enable a torque arising from a pure charge
current bias in Ref. \onlinecite{Hals:epl10} and to compute $\beta$ in Ref. \onlinecite{Hals:prl09}.

\subsubsection{Self-consistency: Spin-battery and enhanced Gilbert Damping}

We discussed two reciprocal effects: torque induced by charge currents
(voltage or electric field) on the magnetization and the current induced by a
time-dependent magnetization. These two effects are not independent. For
instance, in layered systems, when the magnetization precesses, it can pump
spins into adjacent normal metal. The spin-pumping affects magnetization
dynamics depending on whether the spins return into the ferromagnet or not.
When the adjacent normal metal is a good spin sink, this loss of angular
momentum affects the magnetization dynamics by an enhanced Gilbert damping. In
the opposite limit of little or no spin relaxation in an adjacent conductor of
finite size, the pumped steady-state spin-current is canceled by a diffusion
spin current arising from the build-up of spin accumulation potential in the
adjacent conductor. The build-up of the spin accumulation can be interpreted
as a spin battery\cite{Brataas:prb02}. Similarly, in magnetization textures, the dynamic
magnetization pumps currents that in turn exert a torque on the ferromagnet.

In the spin-battery the total spin-current in the normal metal consists of the
diffusion-driven Eq. (\ref{spincurrent}) and the pumped Eq.
(\ref{spin-pumping}) spin currents\cite{Brataas:prb02}. When there are no other intrinsic
time-scales in the transport problem (\textit{e.g.} instantaneous diffusion)
and in the steady state, conservation of angular momentum dictates that the
total spin-current in the normal metal must vanish,%
\[
\mathbf{I}_{N}^{(s,\mathrm{bias})}+\mathbf{I}_{N}^{(s,\mathrm{pump})}=0,
\]
which from Eqs.\ (\ref{spincurrent}) and (\ref{spin-pumping}) results in a
spin accumulation, which can be called a spin-battery bias or spin-motive
force:
\begin{equation}
e\mathbf{V}_{N}^{(s)}=\hbar\mathbf{m}\times\mathbf{\dot{m}}.
\end{equation}
This is a manifestation of Larmor' theorem\cite{Tserkovnyak:rmp05}. In
diffusive systems, the diffusion of the pumped spins into the normal metal
takes a finite amount of time. When the typical diffusion time is longer than
the typical precession time, the AC component averages out to
zero\cite{Brataas:prb02}. In this regime, the spin-battery bias is constant
and determined by
\begin{equation}
\left[  e\mathbf{V}_{N}^{(s)}\right]  ^{(\mathrm{DC})}=\int_{\tau_{\mathrm{p}%
}}\frac{dt}{\tau_{\mathrm{p}}}\mathbf{m}\times\hbar\mathbf{\dot{m}},
\end{equation}
where $\tau_{\mathrm{p}}$ is the precession period. Without spin-flip
processes, the magnitude of the steady-state spin bias is governed by FMR
frequency of the magnetization precession $e\mathbf{V}_{N}^{(s)}=\hbar
\omega_{\mathrm{FMR}}$ and is independent of the interface properties.
Spin-flip scattering in the normal metal reduces the spin bias $e\mathbf{V}%
_{N}^{(s)}<\hbar\omega_{\mathrm{FMR}}$ in a non-universal
way\cite{Brataas:prb02,Tserkovnyak:rmp05}. The loss of spin angular momentum
implies a damping torque on the ferromagnet. Asymmetric spin-flip scattering
rates in adjacent left and right normal metals can also induced a charge
potential difference resulting from the spin-battery, which has been
measured.\cite{Costache:prl06,Wang:prl06} The spin-battery effect has also been measured
via the spin Hall effect in Ref. \cite{Ando:natmat11}.

In the opposite regime, when spins relax much faster than their typical
injection rate into the adjacent normal metal, (\ref{Slonczewski_torqu}), the
net spin-current is well described by the spin-pumping mechanism. According to
Eq. (\ref{spin-pumping}), in which primes may be removed because of the
Onsager reciprocity,
\begin{equation}
{\boldsymbol{\tau}}_{\mathrm{stt}}^{(\mathrm{pump})}=\frac{\gamma\hbar^{2}%
}{2e^2 M_{s}\mathcal{V}}\left[  G_{\perp}^{(R)}\mathbf{m}\times\mathbf{\dot{m}%
}+G_{\perp}^{(I)}\mathbf{\dot{m}}\right]  .\label{torque_pump}%
\end{equation}
We use the superscript \textquotedblleft pump\textquotedblright\ to clarify
that this torque arises from the emission of spins from the ferromagnet. The
first term in Eq.\ (\ref{torque_pump}) is equal to the Gilbert damping term in
the LLG equation (\ref{LLG}). This implies that the spin pumping into an
adjacent conductor maximally enhances the Gilbert damping by
\begin{equation}
\alpha_{\mathrm{stt}}^{(\mathrm{pump})}=\frac{\gamma\hbar^{2}}{2e^2M_{s}%
\mathcal{V}}G_{\perp}^{(R)}.\label{alphaprime}%
\end{equation}
This damping is proportional to the interface conductance $G_{\perp}^{(R)}$
and thus the normal metal-ferromagnet surface area as well as inversely
proportional to the volume of the ferromagnet and therefore scales as
$1/d_{F}$, where $d_{F}$ is the thickness of the ferromagnetic layer. The
transverse conductance per unit areas agrees well with
theory\cite{Tserkovnyak:rmp05}. The microscopic expression for $G_{\perp
}^{(R)}>0$ and therefore $\alpha_{\mathrm{stt}}^{(\mathrm{pump})}>0$. The
second term on the right hand side of Eq.\ (\ref{torque_pump}) in
(\ref{torque_pump}), modifies the gyro-magnetic ratio and $\omega
_{\mathrm{FMR}}$. For conventional ferromagnets like Fe, Ni, and Co,
$G_{\perp}^{(I)}\ll G_{\perp}^{(R)}$ by near cancellation of positive and
negative contributions in momentum space. In these systems $G_{\perp}^{(I)}$ is much smaller than $G_{\perp}^{(R)}$ and the effects of
$G_{\perp}^{(I)}$ might therefore be difficult to observe.

A similar argument leads us to expect an enhancement of the Gilbert damping in
magnetic textures. By inserting the pumped current Eq.
(\ref{currentpumpingcont}) into the torque Eq. (\ref{t_stt}) in place of $\sigma \mathbf{E}$, we find a
contribution caused by the magnetization
dynamics\cite{Foros:prb08,Zhang:prl09,Wong:prb10}%

\begin{equation}
{\boldsymbol{\tau}}_{\mathrm{stt}}^{(\text{\textrm{drift}})}(\mathbf{r}%
)=\frac{\gamma\hbar^{2}}{4e^{2}M_{s}}P^{2}\sigma\left[  \left(  \left[
\mathbf{m\times\partial}_{i}\mathbf{m+}\beta\mathbf{\partial}_{i}%
\mathbf{m}\right]  \cdot\mathbf{\dot{m}}\right)  +\beta\mathbf{m\times}\left(
\left[  \mathbf{m\times\partial}_{i}\mathbf{m+}\beta\mathbf{\partial}%
_{i}\mathbf{m}\right]  \cdot\mathbf{\dot{m}}_{i}\right)  \right]
\mathbf{\partial}_{i}\mathbf{m,}\label{tdrift}%
\end{equation}
which gives rise to additional dissipation of the order $\gamma\hbar^{2}%
P^{2}\sigma/4e^{2}M_{s}\lambda_{w}^{2}$, where $\lambda_{w}$ is the typical
length scale for the variation of the magnetization texture such as the domain
wall width or the radius of a vortex. Eq. (\ref{tdrift}) inserted into the LLG
equation also renormalizes the gyromagnetic ratio by an additional factor
$\beta$. The additional dissipation becomes important for large gradients as
in narrow domain walls and close to magnetic vortex centers
\cite{Foros:prb08,Wong:prb10}.

Finally, we point out that the fluctuation-dissipation theorem dictates that
equilibrium spin-current fluctuations associated with spin-pumping by thermal
fluctuations must lead to magnetization dissipation. This connection was
worked out in Ref. \onlinecite{Foros:prl05}.

\subsection{Onsager Reciprocity Relations\label{Onsager}}

The Onsager reciprocity relations express fundamental symmetries in the linear
response matrix relating thermodynamic forces and currents. In normal metal%
$\vert$%
ferromagnetic heterostructures, a spin accumulation in the normal metal in
contact with a ferromagnet can exert a torque on the ferromagnet, see
Eq.\ (\ref{spintransfertorque}). The reciprocal process is spin pumping, a
precessing ferromagnet induces a spin current in the adjacent normal metal as
described by Eq.\ (\ref{spin-pumping}). Both these effects are non-local since
the spin-transfer torque on the ferromagnet arises from the spin accumulation
potential in the normal metal and the pumped spin current in the normal metal
is a result of the collective magnetization dynamics. In bulk ferromagnets, a
current (or electric field) induces a spin-transfer torque on a magnetization
texture. The reciprocal pumping effect is now an electric current (or
emf)\ generated by the texture dynamics. In the next two subsections we
provide technical details of the derivation of the Onsager reciprocity
relations under these circumstance

\subsubsection{Discrete Systems}

As an example of a discrete system, we consider a normal metal-ferromagnet
bilayer without any spin-orbit interaction (see Ref.
\onlinecite{Hals:epl10}
for a more general treatment that takes spin-flip processes into account) and
under isothermal conditions (the effects of temperature gradients are
discussed in Refs.\ \onlinecite{Hatami:prl07,Kovalev:prb09,Bauer:prb10}). The spin-transfer physics is induced by a pure spin accumulation in the
normal metal, whose creation does not concern us here. The central ingredients
for the Onsager's reciprocity relations are the thermodynamic variables with
associated forces and currents that are related by a linear response
matrix\cite{deGroot:book52}. In order to uniquely define the linear response,
currents $J$ and forces $X$ have to be normalized such that $\dot{F}=\sum XJ.$. This is conventionally done by the
rate of change of the free energy in the non-equilibrium situation in terms of
currents and forces\cite{deGroot:book52}.

Let us consider first the electronic degrees of freedom. In the normal metal
reservoir of a constant spin accumulation $\mathbf{V}_{N}^{(s)}$ the rate of
change of the free energy $F_{N}$ in terms of the total spin 
$\mathbf{s}_{N}$ (in units of electric charge $e$) reads 
\begin{equation}
\dot{F}_{N}=-\mathbf{\dot{s}}_{N}\cdot \mathbf{V}_{N}^{(s)}.
\end{equation}
This identifies $\mathbf{V}_{N}^{(s)}$ as a thermodynamic force that induces
spin currents $\mathbf{I}_{s}=\mathbf{\dot{s}}_{N}$,  which is defined to be positive when leaving
the normal metal. In the ferromagnet, all spins are aligned along the
magnetization direction $\mathbf{m}$. The associated spin accumulation
potential $V_{F}^{(s)}$ can only induce a contribution to the longitudinal
part of the spin current, \textit{e.g.} a contribution to the spin-current
along the magnetization direction $\mathbf{m}$. In our discussion of the
Onsager reciprocity relations, we will set this potential to zero for
simplicity and disregard associated change in the free energy, but it is
straightforward to include the effects of a finite $V_{F}^{(s)}$%
.\cite{Brataas:pr06}

Next, we address the rate of change of the free energy related to the magnetic
degrees of freedom in the ferromagnet,
\[
\dot{F}(\mathbf{m})=-M_{s}\mathcal{V}\mathbf{H}_{\mathrm{eff}}\cdot
\mathbf{\dot{m}}/T,
\]
where $F(\mathbf{m})$ is the magnetic free energy. The total magnetic moment
$M_{s}\mathcal{V}\mathbf{m}$ is a thermodynamic quantity and the effective
magnetic field $\mathbf{H}_{\mathrm{eff}}=-\partial F/\partial(M_{s}%
\mathcal{V}\mathbf{m)}$ is the thermodynamic force that drives the
magnetization dynamics $\mathbf{\dot{m}}$.

In linear response, the spin current $\mathbf{I}_{s}=\mathbf{\dot{s}}$ and
magnetization dynamics $M_{s}\mathcal{V}\mathbf{\dot{m}}$ are related to the
thermodynamic forces as
\begin{equation}
\left(
\begin{array}
[c]{c}%
M_{s}\mathcal{V}\mathbf{\dot{m}}\\
\mathbf{I}_{N}^{(s)}%
\end{array}
\right)  =\left(
\begin{array}
[c]{cc}%
\tilde{L}^{(mm)} & \tilde{L}^{(ms)}\\
\tilde{L}^{(sm)} & \tilde{L}^{(ss)}%
\end{array}
\right)  \left(
\begin{array}
[c]{c}%
\mathbf{H}_{\mathrm{eff}}\\
\mathbf{V}_{N}^{(s)}%
\end{array}
\right)  ,\label{linres_pumpingtorque}%
\end{equation}
where $\tilde{L}^{(mm)}$, $\tilde{L}^{(ms)}$, $\tilde{L}^{(sm)}$, and
$\tilde{L}^{(ss)}$ are $3\times3$ tensors in, \textit{e.g}., a Cartesian basis
for the spin and magnetic moment vectors. Onsager discovered that microscopic
time-reversal symmetry leads to relations between the off-diagonal components
of these linear-response coefficients. Both magnetization in the ferromagnet
and the spin-accumulation in the normal metal are anti-symmetric under
time-reversal leading to the reciprocity relations%
\begin{equation}
L_{ij}^{(sm)}(\mathbf{m})=L_{ji}^{(ms)}(-\mathbf{m}%
).\label{Onsager_pumping_torque}%
\end{equation}
Some care should be taken when identifying the Onsager symmetries in spin
accumulation-induced magnetization dynamics. Specifically, the LLGS equation
(\ref{LLGS}) cannot simply be combined with the linear response relation
(\ref{linres_pumpingtorque}) and Eq. (\ref{Onsager_pumping_torque}). Only the
Landau-Lifshitz-Slonczewski (LL) Eq. (\ref{LL}) directly relates
$\mathbf{\dot{m}}$ to $\mathbf{H}_{\mathrm{eff}}$ as required by
Eq.\ (\ref{linres_pumpingtorque}). In terms of the $3\times3$ matrix
$\tilde{O}$ \textit{e.g.}
\begin{equation}
\tilde{O}_{ij}(\mathbf{m})=\sum_{k}\epsilon_{ikj}m_{k},\label{Oij}%
\end{equation}
where $\epsilon_{ijk}=\frac{1}{2}\left(  j-i\right)  \left(  k-i\right)
\left(  k-j\right)  $ is the Levi-Civita tensor, $\mathbf{m}\times
\mathbf{H}_{\mathrm{eff}}=\tilde{O}\mathbf{H}_{\mathrm{eff}}$, and the LLGS
(\ref{LLGS})\ equation can be written as
\begin{equation}
\left(  1-\alpha\tilde{O}\right)  \mathbf{\dot{m}}=\tilde{O}\left(
-\gamma\mathbf{H}_{\mathrm{eff}}\right)  +{\boldsymbol{\tau}}_{\mathrm{stt}%
}.\label{LLGS_anotherform}%
\end{equation}
By Eq. (\ref{linres_pumpingtorque}), the pumped current in the absence of a
spin accumulation ($\mathbf{V}_{N}^{(s)}=0$) is $\mathbf{I}_{N}^{(s)}%
=\tilde{L}^{(sm)}\mathbf{H}_{\mathrm{eff}}$. Then, by Eq.\ (\ref{spin-pumping}%
), $\mathbf{I}_{N}^{(s)}=\tilde{X}^{(sm)}\mathbf{\dot{m}}$, where the
$3\times3$ tensor $\tilde{X}^{(sm)}$ has components%
\begin{equation}
\tilde{X}_{ij}^{(sm)}(\mathbf{m})=-\frac{\hbar}{e}\left[  G_{\perp}%
^{\prime(R)}\sum_{n}\epsilon_{inj}m_{n}+G_{\perp}^{\prime(I)}\sum
_{nkl}\epsilon_{ink}m_{n}\epsilon_{klj}m_{k}\right]  .
\end{equation}
From the LLG equation (\ref{LLGS_anotherform}) for a vanishing spin accumulation
($\mathbf{V}_{N}^{(s)}=0$) and thus no bias-induced spin-transfer torque
(${\boldsymbol{\tau}}_{\mathrm{stt}}^{\mathrm{(bias)}}=0$), the pumped spin
current can be expressed as $\mathbf{I}_{N}^{(s)}=\tilde{X}^{(sm)}\tilde
{O}\left[  1-\alpha\tilde{O}\right]  ^{-1}\left(  -\gamma\mathbf{H}%
_{\mathrm{eff}}\right)  $, which identifies the linear response coefficient
$\tilde{L}^{(sm)}$ in terms of $\tilde{X}^{(sm)}$ as
\begin{equation}
\tilde{L}^{(sm)}=-\gamma X^{(sm)}\tilde{O}\left[  1-\alpha\tilde{O}\right]
^{-1}.
\end{equation}
Using the Onsager relation (\ref{Onsager_pumping_torque}) and noticing that
$\tilde{O}_{ij}(\mathbf{m})=\tilde{O}_{ji}(-\mathbf{m})$ and $\tilde{X}%
_{ij}^{(sm)}(\mathbf{m})=\tilde{X}_{ji}^{(sm)}(-\mathbf{m})$
\begin{equation}
\tilde{L}^{(ms)}=-\gamma\left[  1-\alpha\tilde{O}\right]  ^{-1}\tilde
{O}X^{(sm)}.
\end{equation}
The rate of change of the magnetization by the spin accumulation therefore
becomes
\begin{align}
\mathbf{\dot{m}}_{\mathrm{stt}} &  =\frac{1}{M_{s}\mathcal{V}}\tilde{L}%
^{(ms)}\mathbf{V}_{N}^{(s)},\nonumber\\
&  =-\frac{\gamma}{M_{s}V}\left[  1-\alpha\tilde{O}\right]  ^{-1}\tilde
{O}X^{(sm)}\mathbf{V}_{N}^{(s)}.
\end{align}
Furthermore, the LLGS equation (\ref{LLGS_anotherform}) in the absence of an
external magnetic field reads $\left[  1-\alpha\tilde{O}\right]
\mathbf{\dot{m}}_{\mathrm{stt}}={\boldsymbol{\tau}}_{\mathrm{stt}%
}^{\mathrm{(drift)}}$. Inserting the phenomenological expression for the
spin-transfer torque (\ref{spintransfertorque}), we identify the linear
response coefficient $\tilde{L}^{(ms)}$:
\begin{align}
{\boldsymbol{\tau}}_{\mathrm{stt}}^{\mathrm{(drift)}} &  =-\frac{\gamma}%
{M_{s}\mathcal{V}}\tilde{O}X^{(sm)}\mathbf{V}_{N}^{(s)}.\nonumber\\
&  =\frac{\gamma}{M_{s}\mathcal{V}e}\left[  G_{\perp}^{\prime(R)}%
\mathbf{m}\times\left(  \mathbf{m}\times\mathbf{V}_{N}^{(s)}\right)
+G_{\perp}^{^{\prime}(I)}\left(  \mathbf{m}\times\mathbf{V}_{N}^{(s)}\right)
\right]  .
\end{align}
This agrees with the phenomenological expression (\ref{spintransfertorque})
when
\begin{equation}
G_{\perp}^{\prime(R)}=G_{\perp}^{(R)};\;G_{\perp}^{^{\prime}(I)}=G_{\perp
}^{(I)}.
\end{equation}
Spin-pumping as expressed by Eq.\ (\ref{spin-pumping}) is thus reciprocal to
the spin-transfer torque as described by Eq.\ (\ref{spintransfertorque}). In
Sec.\ref{mect} these relations are derived by first principles from quantum
mechanical scattering theory, resulting in. $G_{\perp}^{\prime(R)}%
=G_{\uparrow\downarrow}=(e^{2}/h)\sum_{nm}\left[  \delta_{nm}-r_{nm}%
^{\uparrow}\left(  r_{nm}^{\uparrow}\right)  ^{\ast}\right]  $ for a narrow
constriction, where $r_{nm}^{\uparrow}$ ($r_{nm}^{\downarrow}$) is the
reflection coefficient for spin-up (spin-down) electrons from waveguide $m$ to
waveguide mode $n$ . For layered systems with a constant cross section the
microscopic expressions of the transverse (mixing) conductances should be
renormalized by taking into account the contributions from the Sharvin
resistances\cite{Schep:prb97,Bauer:prb03}, which increases the conductance by
roughly a factor of two and is important for a quantitatively comparison
between theory and experiments.\cite{Tserkovnyak:rmp05,Brataas:pr06}

\subsubsection{Continuous Systems}

The Onsager reciprocity relations also relate the magnetization torques and
currents in the magnetization texture of bulk magnets. Following
Refs.\ (\onlinecite{Tserkovnyak:prb08,Hals:prl09}), the rate of change of the
free energy related to the electronic freedom in the ferromagnet is $\dot
{F}_{F}=-\int d\mathbf{r}\dot{q}V$, where $q$ is the charge density and
$eV=\mu$ is the chemical potential. Inserting charge conservation, $\dot
{q}+\nabla\cdot\mathbf{j}=0$ and by partial integration
\begin{equation}
\dot{F}_{F}=-\int dr\mathbf{j\cdot E}\label{entropy_el_cont_current}%
\end{equation}
which identifies charge as a thermodynamic variable, while the electric
field $\mathbf{E}=\mathbf{\nabla}V$ is a thermodynamic force which drives the
current density $\mathbf{j}$. For the magnetic degrees of freedom, the rate of
change of the free energy (or entropy) is
\begin{equation}
\dot{F}_{m}=-M_{s}\int d\mathbf{r\mathbf{\dot{m}}(\mathbf{r})\cdot
H}_{\mathrm{eff}}(\mathbf{r}).
\end{equation}
Just like for discrete systems, $\mathbf{H}_{\mathrm{eff}}(\mathbf{r})$, is
the thermodynamic force and $M_{S}\mathbf{m}$ is the thermodynamic variable to
which it couples. In a local approximation the (linear) response depends only
on the force at the same location:
\begin{equation}
\left(
\begin{array}
[c]{c}%
M_{s}\mathbf{\mathbf{\dot{m}}}\\
\mathbf{j}%
\end{array}
\right)  =\left(
\begin{array}
[c]{cc}%
\tilde{L}^{(mm)} & \tilde{L}^{(mE)}\\
\tilde{L}^{(Em)} & \tilde{L}^{(EE)}%
\end{array}
\right)  \left(
\begin{array}
[c]{c}%
M_{s}\mathbf{H}_{\mathrm{eff}}\\
\mathbf{E}%
\end{array}
\right)  ,
\end{equation}
where $\tilde{L}^{(mm)}$, $\tilde{L}^{(mj)}$, $\tilde{L}^{(jm)}$, and
$\tilde{L}^{(jj)}$ are the local response functions. Onsager's reciprocity
relations dictate again that
\begin{equation}
\tilde{L}_{ji}^{(jm)}(\mathbf{m})=\tilde{L}_{ij}^{(mj)}(-\mathbf{m}%
).\label{Onsager_reciprocity_continous}%
\end{equation}
Starting from the expression for current pumping (\ref{currentpumpingcont}),
we can determine the linear response coefficient $\tilde{L}^{(Em)}$ from
\begin{equation}
\left[  \tilde{L}^{(Em)}\left[  1-\alpha\tilde{O}\right]  \tilde{O}%
^{-1}\right]  _{ij}=-\gamma\frac{\hbar}{2e}\sigma P^{\prime}\left[
\epsilon_{jkl}m_{k}\mathbf{\partial}_{i}m_{l}\mathbf{+}\beta^{\prime
}\mathbf{\partial}_{i}m_{j}\right]  ,
\end{equation}
where the operator $\tilde{O}$ is introduced in the same way as for discrete
systems (\ref{Oij}) to transform the LLG equation into the LL form
(\ref{LLGS_anotherform}). According to Eq.
(\ref{Onsager_reciprocity_continous})
\begin{equation}
\left[  \tilde{O}^{-1}\left[  1-\alpha\tilde{O}\right]  \tilde{L}%
^{(mj)}\right]  _{ij}=-\gamma\frac{\hbar}{2e}\sigma P^{\prime}\left[
\epsilon_{ikl}m_{k}\mathbf{\partial}_{j}m_{l}\mathbf{-}\beta^{\prime
}\mathbf{\partial}_{j}m_{i}\right]  \,.
\end{equation}
The change in the magnetization induced by an electric field is then
$M_{s}\mathbf{\dot{m}}_{\mathrm{stt}}^{(\mathrm{bias})}=\tilde{L}%
^{(mj)}\mathbf{E}$ so that the spin-transfer torque due to a drift current
${\boldsymbol{\tau}}_{\mathrm{stt}}^{\mathrm{(bias)}}=\left[  1-\alpha
\tilde{O}\right]  \mathbf{\dot{m}}_{\mathrm{stt}}^{\mathrm{(bias)}}$ can be
written as
\begin{align}
{\boldsymbol{\tau}}_{\mathrm{stt}}^{\mathrm{(bias)}} &  =-\frac{\gamma\hbar
}{2eM_{s}}\sigma P^{\prime}\epsilon_{imn}m_{m}\left[  \epsilon_{nkl}m_{k}%
E_{j}\mathbf{\partial}_{j}m_{l}\mathbf{-}\beta^{\prime}E_{j}\mathbf{\partial
}_{j}m_{n}\right]  \\
{\boldsymbol{\tau}}_{\mathrm{stt}}^{\mathrm{(bias)}} &  =\gamma\frac{g^{\ast
}\mu_{B}}{2eM_{s}}\sigma P^{\prime}\left[  \left(  \mathbf{E}\cdot
\mathbf{\nabla}\right)  \mathbf{m}+\beta^{\prime}\mathbf{m}\times
\mathbf{E\cdot\nabla m}\right]  \,.
\end{align}
This result agrees with the phenomenological expression for the pumped current
(\ref{currentpumpingcont}) when $P=P^{\prime}$ and $\beta=\beta^{\prime}$.
Therefore, the pumped current and the spin-transfer torque in continuous
systems are reciprocal processes. The pumped current can be formulated as the
response to a spin-motive force\cite{Barnes:prl07}.

In small systems and thin wires, the current-voltage relation is not well
represented by a local approximation. A global approach based on collective
coordinates as outlined around Eq. (\ref{Collective_coordinate}) is then a
good choice to keep the computational effort in check. Of course, the Onsager
reciprocity relations between the pumped current and the effective
current-induced torques on the magnetization hold then as
well\cite{Hals:prl09}.

\section{Microscopic Derivations}

\subsection{Spin-transfer Torque}

\subsubsection{Discrete Systems - Magneto-electronic Circuit
Theory\label{mect}}

\begin{figure}[ptb]
\includegraphics[width=0.5\columnwidth]{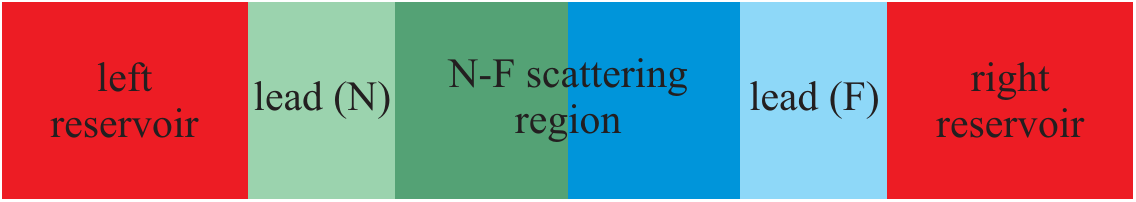} \caption{Schematic
of how transport between a normal metal and a ferromagnet is computed by
scattering theory. The scattering region, which may contain the normal
metal-ferromagnet interface and diffusive parts of the normal metal as well as
ferromagnet, is attached to real or fictious leads that are in contact with a
left and right reservoir. In the reservoirs, the distributions of charges and
spins are assumed to be known via the charge potential and spin accumulation
bias. }%
\label{fig:scatteringtheory}%
\end{figure}

Physical properties across a scattering region can be expressed in terms of
the region's scattering matrix, which requires a separation of the system into
reservoirs, leads, and a scattering region, see
Fig.\ (\ref{fig:scatteringtheory}). In the lead with index $\alpha$, the field operator for spin
$s$-electrons is\cite{Buttiker:prb92}
\begin{equation}
\hat{\Psi}_{\alpha}^{(s)}=\int\frac{d\epsilon}{\sqrt{2\pi}}\left[  v_{\alpha
}^{(ns)}\right]  ^{-1/2}\sum_{n\sigma}\varphi_{\alpha}^{(ns)}%
(\boldsymbol{\varrho})e^{-i\epsilon_{\alpha}^{(nks)}t/\hbar}\left[
e^{ikx}\hat{a}_{\alpha}^{(ns)}(\epsilon)+e^{-ikx}\hat{b}_{\alpha}%
^{(ns)}(\epsilon)\right]  \label{fieldoperator}%
\end{equation}
in terms of the annihilation operators $\hat{a}_{\alpha}^{(ns)}$ ($b_{\alpha
}^{(ns)}$)\ for particles incident on (outgoing from) the scattering region
in transverse wave guide modes with orbital quantum number $n$ and spin
quantum number $s$ ($s=\uparrow$ or $s=\downarrow$). Furthermore, the
transverse wave function is $\varphi_{\alpha}^{(ns)}(\boldsymbol{\varrho})$,
the transverse coordinate $\boldsymbol{\varrho}$, the longitudinal coordinate
along the waveguide is $x$ and $v_{\alpha}^{(ns)}$ is the longitudinal velocity
for waveguide mode $ns$. The positive definite momentum $k$ is related to the
energy $\epsilon$ by $\hbar k=(2m\epsilon)^{1/2}$. The annihilation operators
for incident and outgoing electrons are related by the scattering matrix
\begin{equation}
\hat{b}_{\alpha}^{(ns)}(\epsilon)=\sum_{\beta ms^{\prime}}S_{\alpha\beta
}^{(nsms^{\prime})}(\epsilon)\hat{a}_{\beta}^{(ms^{\prime})}(\epsilon).
\label{scatteringmatrix}%
\end{equation}
In the basis of the leads ($\alpha=N$ (normal metal) or $\alpha=F$
(ferromagnet)), the scattering matrix is
\[
S=\left(
\begin{array}
[c]{cc}%
r & t\\
t^{\prime} & r^{\prime}%
\end{array}
\right)  \, ,
\]
where $r$ ($t$) is a matrix of the reflection (transmission) coefficients
between the wave guide modes for an electron incident from the left.
Similarly, $r^{\prime}$ and $t^{\prime}$ characterize processes where the
electron is incident from the right.

In terms of the field operators defined by Eq. (\ref{fieldoperator}) and the
scattering matrix Eq. (\ref{scatteringmatrix}), at low frequencies, the spin
current that flows in the normal metal $\alpha=N$ in the direction towards the
scattering region is
\begin{equation}
\mathbf{I}_{\alpha}^{(s)}(t)=\frac{e}{h}\int_{-\infty}^{\infty}d\epsilon
_{1}\int_{-\infty}^{\infty}d\epsilon_{2}{\displaystyle\sum\limits_{\beta
\gamma}}\sum_{nml}{\displaystyle\sum\limits_{\sigma\sigma^{\prime}}}%
\exp(i\left(  \epsilon_{1}-\epsilon_{2}\right)  t/\hbar)\mathbf{A}%
_{\alpha\beta,\alpha\gamma}^{(nm,nl),(\sigma,\sigma^{\prime})}(\epsilon
_{1},\epsilon_{2})\hat{a}_{\beta}^{(m\sigma)\dag}(\epsilon_{1})\hat{a}%
_{\gamma}^{(l\sigma^{\prime})}(\epsilon_{2}),\label{scattering_spincurrent}%
\end{equation}
where
\[
\mathbf{A}_{\alpha\beta,\alpha\gamma}^{(nm,nl)(\sigma,\sigma^{\prime}%
)}(\epsilon_{1},\epsilon_{2})={\displaystyle\sum\limits_{ss^{\prime}}}\left[
\delta_{\alpha\beta}\delta^{(nm)}\delta^{(s\sigma)}\delta_{\alpha\gamma}%
\delta^{(nl)}\delta^{(s^{\prime}\sigma^{\prime})}-S_{\alpha\beta}%
^{(ns,m\sigma)\ast}(\epsilon_{1})S_{\alpha\gamma}^{(ns^{\prime},l\sigma
^{\prime})}(\epsilon_{2})\right]  {\boldsymbol{\sigma}}^{(ss^{\prime})}%
\]
and $\boldsymbol{\sigma}^{(ss^{\prime})}$ is a vector of the $2\times2$ Pauli
matrices that depends on the spin indices $s$ and $s^{\prime}$ of the
waveguide mode. The charge current can be found in a similar way. We are
interested in the expectation value of the spin-current
(\ref{scattering_spincurrent}) when the system is driven out-of-equilibrium.
In \textit{equilibrium}, the expectation values are
\begin{equation}
\left\langle \hat{a}_{\alpha}^{(ns)\dag}(\epsilon)\hat{a}_{\beta}%
^{(ms^{\prime})}(\epsilon^{\prime})\right\rangle _{\mathrm{eq}}=\delta
(\epsilon-\epsilon^{\prime})\delta_{\alpha\beta}\delta^{(ss^{\prime})}%
\delta^{(nm)}f_{\mathrm{FD}}(\epsilon),\label{exp_localeq}%
\end{equation}
where $f_{\mathrm{FD}}(\epsilon)$ is the Fermi-Dirac distribution of electrons
with energy $\epsilon$. A non-equilibrium spin-accumulation in the normal
metal reservoir is not captured by the local equilibrium ansatz in
Eq.\ (\ref{exp_localeq}), however. A spin accumulation in the normal metal
reservoir can still be postulated when spin-flip dissipation is slow compared
to all other relevant time scales. We assume the normal metal and ferromagnet
have an isotropic distribution of spins in the orbital space, and for clarity consider no charge bias. The expectation
for the number of charges and spins in the waveguide describing normal metal
leads attached to the normal reservoirs are
\begin{equation}
\left\langle \hat{a}_{N}^{(ns)\dag}(\epsilon)\hat{a}_{N}^{(ms^{\prime}%
)}(\epsilon^{\prime})\right\rangle =\delta(\epsilon-\epsilon^{\prime})\left[
\delta^{(mn)}\delta^{(ss^{\prime})}f_{\mathrm{FD}}(\epsilon)+\delta
^{(mn)}f_{N}^{(s^{\prime}s)}(\epsilon)\right]  \, .
\end{equation}
The spin-accumulation $\mathbf{V}_{N}^{(s)}$ is related to the $2\times2$
out-of-equilibrium distribution matrix $f_{N}^{(s^{\prime}s)}(\epsilon)$ by
\begin{equation}
\boldsymbol{\sigma}^{(ss^{\prime})}\cdot\mathbf{V}_{N}^{(s)}=\int_{-\infty
}^{\infty}d\epsilon f_{N}^{(ss^{\prime})}(\epsilon)/e \, .
\label{spinaccumulation}%
\end{equation}
For the spin-transfer physics, a bias voltage in the ferromagnet does not
contribute since it only gives rise to a charge current and a longitudinal
spin current. As in the previous section, we therefore set this voltage to
zero for simplicity, so that in the ferromagnetic lead attached to the
ferromagnetic reservoir
\begin{equation}
\left\langle \hat{a}_{F}^{(ns)\dag}(\epsilon)\hat{a}_{F}^{(ms^{\prime}%
)}(\epsilon^{\prime})\right\rangle =\delta(\epsilon-\epsilon^{\prime}%
)\delta^{(ms)}\delta^{(s^{\prime}s)}f_{\mathrm{FD}}(\epsilon).
\end{equation}
Furthermore, the expectation values of the cross-correlations remain zero also
out-of-equilibrium, $\left\langle \hat{a}_{N}^{(ns)\dag}(\epsilon)\hat{a}%
_{F}^{(ms^{\prime})}(\epsilon^{\prime})\right\rangle =0$. The spin current in
lead $\alpha$ is then%
\begin{equation}
\mathbf{I}_{\alpha}^{(s)}(t)=\frac{e}{h}\int_{-\infty}^{\infty}d\epsilon%
{\displaystyle\sum\limits_{nml}}
{\displaystyle\sum\limits_{ss^{\prime}\sigma\sigma^{\prime}}}
\left[  \delta^{(nm)}\delta^{(s\sigma)}\delta^{(nl)}\delta^{(s^{\prime}%
\sigma^{\prime})}-r_{NN}^{(ns,m\sigma)\ast}r_{NN}^{(ns^{\prime},l\sigma
^{\prime})}\right]  \boldsymbol{\sigma}^{(\sigma\sigma^{\prime})}%
f^{(\sigma^{\prime}\sigma)}.
\end{equation}
Without spin-flip scattering, the reflection coefficient can be expressed as%
\begin{equation}
r_{NN}^{nsm\sigma}=\left(  r_{NN}^{nm,\uparrow}+r_{NN}^{nm,\downarrow}\right)
\delta^{(s\sigma)}/2+\mathbf{m}\cdot\boldsymbol{\sigma}_{s\sigma}\left(
r_{NN}^{nm,\uparrow}-r_{NN}^{nm,\downarrow}\right)  /2
\end{equation}
which can be represented in spin space as
\begin{equation}
r_{NN}^{nsm\sigma}=r_{NN}^{nm,(c)}1+r_{NN}^{nm,(s)}\mathbf{m}\cdot
\boldsymbol{\sigma}%
\end{equation}
since the scattering matrix can be decomposed into components aligned and
anti-aligned with the magnetization direction. These matrices only depend on
the orbital quantum numbers ($n$ and $m$). Using the representation of the
out-of-equilibrium spin density in terms of the spin accumulation
(\ref{spinaccumulation}) \cite{Brataas:prl00},
\begin{equation}
\mathbf{I}_{N}^{(s)}=\left(  G_{\uparrow}+G_{\downarrow
}\right)  \mathbf{m}\left(  \mathbf{m}\cdot\mathbf{V}_{N}^{(s)}\right)
\mathbb{-}2G_{\perp}^{(R)}\mathbf{m}\times\left(  \mathbf{m}\times
\mathbf{V}_{N}^{(s)}\right)  -2G_{\perp}^{(I)}\left(  \mathbf{m}%
\times\mathbf{V}_{N}^{(s)}\right)  \label{SM-spincurrent}%
\end{equation}
in agreement with (\ref{spincurrent}) when there is no bias voltage in the
ferromagnet ($V_{F}=0$) which we have assumed for clarity here. We identify the microscopic expressions for the
conductances\cite{Brataas:prl00} associated with spins aligned and
anti-aligned with the magnetization direction
\begin{align}
G_{\uparrow} &  =\frac{e^{2}}{h}%
\sum_{nm}
\left[  \delta_{nm}-\left\vert r_{NN}^{nm,\uparrow}\right\vert^2 \right]  ,\\
G_{\uparrow} &  =\frac{e^{2}}{h}%
\sum_{nm}
\left[  \delta_{nm}-\left\vert r_{NN}^{nm,\uparrow}\right\vert^2 \right]  ,
\end{align}
and the transverse (complex valued) spin-mixing conductance%
\begin{equation}
G_{\perp}=\frac{e^{2}}{h}%
\sum_{n_m}
\left[  \delta_{nm}-r_{NN}^{nm,\uparrow}r_{NN}^{nm,\downarrow\ast}\right]  .
\end{equation}
These results are valid when the transmission coefficients are small such that
currents do not affect the reservoirs. Otherwise, the transverse conductance
parameters should be renormalized by taking into account the Sharvin
resistances, as described above\cite{Schep:prb97,Bauer:prb03}. In the limit we
considered here, the expression for the spin-current depends only on the
reflection coefficients for transport from the normal metal towards the
ferromagnet and not on the transmission coefficients for propagation from the
normal metal into the ferromagnet. This follows from our assumption that the
ferromagnet is longer than the transverse coherence length as well as our
disregard of the spin accumulation in the ferromagnet. Both assumptions can be
easily relaxed if necessary\cite{Tserkovnyak:rmp05,Brataas:pr06}.

\subsubsection{Continuous Systems}

\label{C-STT}

Spin torques in continuous spin textures can be studied by either quantum
kinetic theory,\cite{Tserkovnyak:prb06} imaginary-time\cite{Kohno:jopj06} and
functional Keldysh\cite{duinePRB07fk} diagrammatic approaches, or the
scattering-matrix formalism.\cite{Hals:prl09} The latter is particularly
powerful when dealing with nontrivial band structures with strong spin-orbit
interactions, while the others give complementary insight, but are mostly
limited to simple model studies. When the magnetic texture is sufficiently
smooth on the relevant length scales (the transverse spin coherence length
and, in special cases, the spin-orbit precession length) the spin torque can
be expanded in terms of the local magnetization and current density as well as
their spatial-temporal derivatives. An example is the phenomenological Eq.
(\ref{t_stt}) for the electric-field driven magnetization dynamics of an
isotropic ferromagnet. While the physical meaning of the coefficients is
clear, the microscopic origin and magnitude of the dimensionless parameter
$\beta$ has still to be clarified.

The solution of the LLG equation (\ref{LLG}) appended by these spin torques
depends sensitively on the relationship between the dimensionless Gilbert
damping constant $\alpha$ and the dissipative spin-torque parameter $\beta$:
the special case $\beta/\alpha=1$ effectively manifests Galilean
invariance\cite{barnesPRL05} while the limits $\beta/\alpha\gg1$ and
$\beta/\alpha\ll1$ are regimes of qualitatively distinct macroscopic behavior.
The ratio $\beta/\alpha$ determines the onset of the ferromagnetic
current-driven instability\cite{Tserkovnyak:prb06} as well as the Walker
threshold\cite{schryerJAP74} for the current-driven domain-wall
motion\cite{Thiaville:epl05}, and both diverge as $\beta/\alpha\rightarrow1$.
The sub-threshold current-driven domain-wall velocity is proportional to
$\beta/\alpha$,\cite{Zhang:prl04} while $\beta/\alpha=1$ in a special point,
at which the effect of a uniform current density $\mathbf{j}$ on the
magnetization dynamics is eliminated in the frame of reference that moves with
velocity $\mathbf{v}\propto\mathbf{j}$, which is of the order of the electron
drift velocity.\cite{tserkovJMMM08} Although the exact ratio $\beta/\alpha$ is
a system-dependent quantity, some qualitative aspects not too sensitive to the
microscopic origin of these parameters have been discussed in relation to
metallic
systems.\cite{barnesPRL05,Tserkovnyak:prb06,Kohno:jopj06,skadsemPRB07}
However, these approaches fail for strongly spin-orbit coupled systems such as
dilute magnetic semiconductors \cite{Hals:prl09}.

Let us outline the microscopic origin of $\beta$ for a simple toy model for a
ferromagnet. In Ref.~\onlinecite{Tserkovnyak:prb06}, we developed a
self-consistent mean-field approach, in which itinerant electrons are
described by a single-particle Hamiltonian
\begin{equation}
\mathcal{\hat{H}}=\left[  \mathcal{H}_{0}+U(\mathbf{r},t)\right]  \hat
{1}+\frac{\gamma\hbar}{2}\boldsymbol{\hat{\sigma}}\cdot\left(  \mathbf{H}%
+\mathbf{H}_{\mathrm{xc}}\right)  (\mathbf{r},t)+\mathcal{\hat{H}}_{\sigma}\,,
\label{HKS}%
\end{equation}
where the unit matrix $\hat{1}$ and a vector of Pauli matrices
$\boldsymbol{\hat{\sigma}}=(\hat{\sigma}_{x},\hat{\sigma}_{y},\hat{\sigma}%
_{z})$ form a basis for the Hamiltonian in spin space. $\mathcal{H}_{0}$ is
the crystal Hamiltonian including kinetic and potential energy. $U$ is the
scalar potential consisting of disorder and applied electric-field
contributions. The total magnetic field consists of the applied, $\mathbf{H}$,
and exchange, $\mathbf{H}_{\mathrm{xc}}$, fields that, like $U$, are
parametrically time dependent. Finally, the last term in the Hamiltonian,
$\mathcal{\hat{H}}_{\sigma}$, accounts for spin-dephasing processes,
\textit{e.g}, due to quenched magnetic disorder or spin-orbit scattering
associated with impurity potentials. This last term is responsible for
low-frequency dissipative processes affecting dimensionless parameters
$\alpha$ and $\beta$ in the collective equation of motion.

In the time-dependent spin-density-functional theory
\cite{rungePRL84,capellePRL01,qianPRL02} of itinerant ferromagnetism, the
exchange field $\mathbf{H}_{\mathrm{xc}}$ is a functional of the
time-dependent spin-density matrix
\begin{equation}
\rho_{\alpha\beta}(\mathbf{r},t)=\langle\hat{\Psi}_{\beta}^{\dagger
}(\mathbf{r})\hat{\Psi}_{\alpha}(\mathbf{r})\rangle_{t}\,,
\end{equation}
where $\hat{\Psi}$'s are electronic field operators, which should be computed
self-consistently as solutions of the Schr\"{o}dinger equation for
$\mathcal{\hat{H}}$. The spin density of conducting electrons is given by
\begin{equation}
\mathbf{s}(\mathbf{r})=\frac{\hbar}{2}\mbox{Tr}\left[  \boldsymbol{\hat
{\sigma}}\hat{\rho}(\mathbf{r})\right]  \,.\label{s}%
\end{equation}
We focus on low-energy magnetic fluctuations that are long ranged and
transverse and restrict our attention to a single parabolic band.
Consideration of more realistic band structures is also in principle possible
from this starting point\cite{Garate:prb09}. We adopt the adiabatic
local-density approximation (ALDA, essentially the Stoner model) for the
exchange field:
\begin{equation}
\gamma\hbar\mathbf{H}_{\mathrm{xc}}[\hat{\rho}](\mathbf{r},t)\approx
\Delta_{\mathrm{xc}}\mathbf{m}(\mathbf{r},t)\,,\label{Hxc}%
\end{equation}
with direction $\mathbf{m}=-\mathbf{s}/s$ locked to the time-dependent spin
density (\ref{s}).

In another simple model of ferromagnetism, the so-called $s$-$d$ model,
conducting $s$ electrons interact with the exchange field of the $d$ electrons
that are assumed to be localized to the crystal lattice sites. The $d$-orbital
electron spins account for most of the magnetic moment. Because $d$-electron
shells have large net spins and strong ferromagnetic correlations, they are
usually treated classically. In a mean-field $s$-$d$ description, therefore,
conducting $s$ orbitals are described by the same Hamiltonian (\ref{HKS}) with
an exchange field (\ref{Hxc}). The differences between the Stoner and $s$-$d$
models for the magnetization dynamics are subtle and rather minor. In the
ALDA/Stoner model, the exchange potential is (on the scale of the
magnetization dynamics) instantaneously aligned with the total magnetization.
In contrast, the direction of the unit vector $\mathbf{m}$ in the $s$-$d$
model corresponds to the $d$ magnetization, which is allowed to be slightly
misaligned with the $s$ magnetization, transferring angular momentum between
the $s$ and $d$ magnetic moments. Since most of the magnetization is carried
by the latter, the external field $\mathbf{H}$ couples mainly to the $d$
spins, while the $s$ spins respond to and follow the time-dependent exchange
field (\ref{Hxc}). As $\Delta_{\mathrm{xc}}$ is usually much larger than the
external (including demagnetization and anisotropy) fields that drive
collective magnetization dynamics, the total magnetic moment will always be
very close to $\mathbf{m}$. A more important difference of the philosophy
behind the two models is the presumed shielding of the $d$ orbitals from
external disorder. The reduced coupling with dissipative degrees of freedom
would imply that their dynamics are more coherent. Consequently, the
magnetization damping has to originate from the disorder experienced by the
itinerant $s$ electrons. As in the case of the itinerant ferromagnets, the
susceptibility has to be calculated self-consistently with the magnetization
dynamics parametrized by $\mathbf{m}$. For more details on this model, we
refer to Refs.\ \onlinecite{tserkovAPL04} and \onlinecite{Tserkovnyak:prb06}.
With the above differences in mind, the following discussion is applicable to
both models. The Stoner model is more appropriate for transition-metal
ferromagnets because of the strong hybridization between $d$ and $s,p$
electrons. For dilute magnetic semiconductors with by deep magnetic impurity
states the $s$-$d$ model appears to be a better choice.

The single-particle itinerant electron response to electric and magnetic
fields in Hamiltonian (\ref{HKS}) is all that is needed to compute the
magnetization dynamics microscopically. Stoner and $s$-$d$ models have to be
distinguished only at the final stages of the calculation, when we
self-consistently relate $\mathbf{m}(\mathbf{r},t)$ to the electron spin
response. The final result for the simplest parabolic-band Stoner model with
isotropic spin-flip disorder comes down to the torque (\ref{t_stt}) with
$\alpha\approx\beta$. The latter is proportional to the spin-dephasing rate
$\tau_{\sigma}^{-1}$ of the itinerant electrons:
\begin{equation}
\beta\approx\frac{\hbar}{\tau_{\sigma}\Delta_{\mathrm{xc}}}\,. \label{Ybeta}%
\end{equation}
The derivation assumes $\omega,\tau_{\sigma}^{-1}\ll\Delta_{\mathrm{xc}}%
/\hbar$, which is typically the case in real materials sufficiently below the
Curie temperature. The $s$-$d$ model yields the same result for $\beta$,
Eq.~(\ref{Ybeta}), but the Gilbert damping constant
\begin{equation}
\alpha\approx\eta\beta\label{ab}%
\end{equation}
is reduced by the ratio $\eta$ of the itinerant to the total angular momentum
when the $d$-electron spin dynamics is not damped. [Note that Eq.~(\ref{ab})
is also valid for the Stoner model since then $\eta=1$.]

These simple model considerations shed light on the microscopic origins of
dissipation in metallic ferromagnet as reflected in the $\alpha$ and $\beta$
parameters. In Sec.~\ref{abinitio} we present a more systematic,
first-principle approach based on the scattering-matrix approach, which
accesses the material dependence of both $\alpha$ and $\beta$ with realistic
electronic band structures.

\subsection{Spin Pumping}

\subsubsection{Discrete Systems}

When the scattering matrix is time-dependent, the energy of outgoing and
incoming states does not have to be conserved and the scattering relation
(\ref{scatteringmatrix}) needs to be appropriately
generalized\cite{buttikerZPB94}. We will demonstrate here how this is done in
the limit of slow magnetization dynamics, \textit{i.e.}, adiabatic pumping.
When the time dependence of the scattering matrix $\hat{S}_{\alpha\beta
}^{(nm)}[X_{i}(t)]$ is parameterized by a set of real-valued parameters
$X_{i}(t)$, the pumped spin current in excess of its static bias-driven value
(\ref{SM-spincurrent}) is given by\cite{Tserkovnyak:prl02}
\begin{equation}
\mathbf{I}_{\alpha}^{s}(t)= e\sum_{i}\frac{\partial
\mathbf{n}_{\alpha}}{\partial X_{i}}\frac{dX_{i}(t)}{dt}\,,\label{SM-pumping}%
\end{equation}
where the \textquotedblleft spin emissivity\textquotedblright\ vector\ by the
scatterer into lead $\alpha$ is\cite{brouwerPRB98}
\begin{equation}
\frac{\partial\mathbf{n}_{\alpha}}{\partial X_{i}}=\frac{1}{2\pi}%
\mathrm{Im}\sum_{\beta}\sum_{mn}\sum_{ss^{\prime}\sigma}\frac{\partial
S_{\alpha\beta}^{(ms,n\sigma)\ast}}{\partial X_{i}}\boldsymbol{\hat{\sigma}%
}^{(ss^{\prime})}S_{\alpha\beta}^{(ms^{\prime},n\sigma)}%
\,.\label{SM-emissivity}%
\end{equation}
Here, $\boldsymbol{\hat{\sigma}}^{(ss^{\prime})}$ is again the vector of Pauli
matrices. In the case of a magnetic monodomain insertion and in the absence of
spin-orbit interactions, the spin-dependent scattering matrix between the
normal-metal leads can be written in terms of the respective spin-up and
spin-down scattering matrices:\cite{Brataas:prl00}
\begin{equation}
S_{\alpha\beta}^{(ms,ns^{\prime})}[\mathbf{m}]=\frac{1}{2}S_{\alpha\beta
}^{(mn)\uparrow}\left(  \delta^{(ss^{\prime})}+\mathbf{m}\cdot\boldsymbol{\hat
{\sigma}}^{(ss^{\prime})}\right)  +\frac{1}{2}S_{\alpha\beta}^{(mn)\downarrow
}\left(  \delta^{(ss^{\prime})}-\mathbf{m}\cdot\boldsymbol{\hat{\sigma}%
}^{(ss^{\prime})}\right)  .\label{SM-projected}%
\end{equation}
Here, $\mathbf{m}(t)$ is the unit vector along the magnetization direction and
$\uparrow$ ($\downarrow$) are spin orientations defined along (opposite) to
$\mathbf{m}$.

Spin pumping due to magnetization dynamics $\mathbf{m}(t)$ is then found by
substituting Eq.~(\ref{SM-projected}) into Eqs.~(\ref{SM-emissivity}) and
(\ref{SM-pumping}). After straightforward algebra:\cite{Tserkovnyak:prl02}
\begin{equation}
\mathbf{I}_{\alpha}^{s}(t)=\left(  \frac{\hbar}{e}\right)
\left(  G_{\perp}^{(R)}\mathbf{m}\times\frac{d\mathbf{m}}{dt}+G_{\perp
}^{(I)}\frac{d\mathbf{m}}{dt}\right)  \,. \label{SM-sp}%
\end{equation}
As before, we assume here a sufficiently thick ferromagnet, on the scale of
the transverse spin-coherence length. Note that the spin pumping is expressed
in terms of the same complex-valued mixing conductance $G_{\perp}=G_{\perp
}^{(R)}+iG_{\perp}^{(I)}$ as the dc current (\ref{SM-spincurrent}), in
agreement with the Onsager reciprocity principle as found on phenomenological
grounds in Sec. \ref{Onsager}.

Charge pumping is governed by expressions similar to Eqs.~(\ref{SM-pumping})
and (\ref{SM-emissivity}), subject to the following substitution: $\boldsymbol{\hat{\sigma}%
}\rightarrow\delta$ (Kronecker delta). A finite charge pumping by a monodomain
magnetization dynamics into normal-metal leads, however, requires a
ferromagnetic analyzer or finite spin-orbit interactions and appropriately
reduced symmetries, as discussed in Refs.~\onlinecite{chernyshovNATP09,manchonPRB08,garatePRB09,Hals:epl10}.

An immediate consequence of the pumped spin current (\ref{SM-sp}) is an
enhanced Gilbert damping of the magnetization
dynamics.\cite{Tserkovnyak:prl02} Indeed, when the reservoirs are good spin
sinks and spin backflow can be disregarded, the spin torque associated with
the spin current (\ref{SM-sp}) into the $\alpha$-th lead, as dictated by the
conservation of the spin angular momentum, Eq.~(\ref{Slonczewski_torqu}),
contributes (\textit{cf}. Eq. (\ref{alphaprime})):
\begin{equation}
\alpha^{\prime}=g^{\ast}\frac{\hbar\mu_{B}}{2e^{2}}\frac{G_{\perp}^{(R)}%
}{M_{s}\mathcal{V}}\label{SM-damping}%
\end{equation}
to the Gilbert damping of the ferromagnet in Eq.~(\ref{LLG}). Here, $g^{\ast
}\sim2$ is the $g$ factor of the ferromagnet, $M_{s}\mathcal{V}$ its total
magnetic moment, and $\mu_{B}$ is Bohr magneton. For simplicity, we neglected
$G_{\perp}^{(I)}$, which is usually not important for inter-metallic
interfaces. If we disregard energy relaxation processes inside the
ferromagnet, which would drain the associated energy dissipation out of the
electronic system, the enhanced energy dissipation associated with the Gilbert
damping is associated with heat flows into the reservoirs. Phenomenologically,
the dissipation power follows from the magnetic free energy $F$ and the LLG
Eq. (\ref{LLG}) as
\begin{equation}
P\equiv-\partial_{\mathbf{m}}F_{m}\cdot \dot{\mathbf{m}}=M_{s}\mathcal{V}%
\mathbf{H}_{\mathrm{eff}}\cdot\dot{\mathbf{m}}=\frac{\alpha M_{s}\mathcal{V}%
}{\gamma}\dot{\mathbf{m}}^{2}%
\end{equation}
or, more generally, for anisotropic damping (with, for simplicity, an
isotropic gyromagnetic ratio), by
\begin{equation}
P=\frac{M_{s}\mathcal{V}}{\gamma}\dot{\mathbf{m}}\cdot\tensor{\alpha}\cdot
\dot{\mathbf{m}}\,.
\end{equation}

Heat flows can be also calculated microscopically by the scattering-matrix
transport formalism. At low temperatures, the heat pumping rate into the
$\alpha$-th lead is given
by\cite{avronPRL01,moskaletsPRB02dn,moskaletsPRB02fl}
\begin{equation}
I_{\alpha}^{E}=\frac{\hbar}{4\pi}\sum_{\beta}\sum_{mn}\sum_{ss^{\prime}%
}\left\vert \dot{S}_{\alpha\beta}^{(ms,ns^{\prime})}\right\vert ^{2}%
=\frac{\hbar}{4\pi}\sum_{\beta}\mathrm{Tr}\left(  \hat{\dot{S}}_{\alpha\beta
}^{\dagger}\hat{\dot{S}}_{\alpha\beta}\right)  \,,\label{pump-IE}%
\end{equation}
where the carets denote scattering matrices with suppressed transverse-channel
indices. When the time dependence is entirely due to the magnetization
dynamics, $\dot{S}_{\alpha\beta}^{(ms,ns^{\prime})}=\partial_{\mathbf{m}%
}S_{\alpha\beta}^{(ms,ns^{\prime})}\cdot\dot{\mathbf{m}}$. Utilizing again
Eq.~(\ref{SM-projected}), we find for the heat current into the $\alpha$-th
lead:\cite{Brataas:prl08}
\begin{equation}
I_{\alpha}^{E}=\dot{\mathbf{m}}\cdot\tensor{G}_{\alpha}\cdot\dot{\mathbf{m}%
}\,,
\end{equation}
in terms of the dissipation tensor\cite{Brataas:prl08}
\begin{equation}
G_{\alpha}^{ij}=\frac{\gamma^2 \hbar}{4\pi}\mathrm{Re}\sum_{\beta}\mathrm{Tr}\left(
\frac{\partial\hat{S}_{\alpha\beta}^{\dagger}}{\partial m_{i}}\frac
{\partial\hat{S}_{\alpha\beta}}{\partial m_{j}}\right)
\label{Gilbertscattering}
\end{equation}
In the limit of vanishing spin-flip in the ferromagnet, meaning that all
dissipation takes place in the reservoirs, we find
\begin{equation}
G_{\alpha}^{ij}=\frac{\gamma^2 \hbar}{4\pi}\mathrm{Re}\sum_{\beta}\mathrm{Tr}\left(
\frac{\partial\hat{S}_{\alpha\beta}^{\dagger}}{\partial m_{i}}\frac
{\partial\hat{S}_{\alpha\beta}}{\partial m_{j}}\right)  =\gamma^2 \frac{1}{2}\left(
\frac{\hbar}{e}\right)  ^{2}G_{\perp}^{(R)}\delta_{ij}\,.\label{SM-Gij}%
\end{equation}
Equating this $I_{\alpha}^{E}$ with $P$ above, we obtain a microscopic
expression for the Gilbert damping tensor $\tensor{\alpha}$:
\begin{equation}
\tensor{\alpha}=g^{\ast}\frac{\hbar\mu_{B}}{2e^{2}}\frac{G_{\perp}^{(R)}%
}{M_{s}\mathcal{V}}\tensor{1},\label{SM-energy}%
\end{equation}
which agrees with Eq.~(\ref{SM-damping}). Indeed, in the absence of spin-orbit
coupling the damping is necessarily isotropic. While Eq.~(\ref{SM-damping})
reproduces the additional Gilbert damping due to the interfacial spin pumping,
Eq.~(\ref{SM-Gij}) is more general, and can be used to compute bulk
magnetization damping, as long as it is of a purely electronic
origin\cite{Brataas:prl08,Starikov:prl10}.

\subsubsection{Continuous Systems}

As has already been noted, spin pumping in continuous systems is the Onsager
counterpart of the spin-transfer torque discussed in Sec.~\ref{C-STT}%
.\cite{Tserkovnyak:prb08} While a direct diagrammatic calculation for this
pumping is possible\cite{Duine:prb08}, with results equivalent to those of the
quantum-kinetic description of the spin-transfer torque outlined above, we
believe that the scattering-matrix formalism is the most powerful microscopic
approach\cite{Hals:prl09}. The latter is particularly suitable for
implementing parameter-free computational schemes that allow a realistic
description of material-dependent properties.

An important example is pumping by a moving domain wall in a
quasi-one-dimensional ferromagnetic wire. When the domain wall is driven by a
weak magnetic field, its shape remains to a good approximation unaffected, and
only its position $r_{w}(t)$ along the wire is needed to parameterize its slow
dynamics. The electric current pumped by the sliding domain wall into the
$\alpha$-th lead can then be viewed as pumping by the $r_{w}$ parameter, which
leads to\cite{brouwerPRB98}
\begin{equation}
I_{\alpha}^{c}=\frac{e\dot{r}_{w}}{2\pi}\mathrm{Im}\sum_{\beta}\mathrm{Tr}%
\left(  \frac{\partial\hat{S}_{\alpha\beta}}{\partial r_{w}}\hat{S}%
_{\alpha\beta}^{\dagger}\right)  \,. \label{CP-Ic}%
\end{equation}
The total heat flow into both leads induced by this dynamics is according to
Eq.~(\ref{pump-IE})
\begin{equation}
I^{E}=\frac{\hbar\dot{r}_{w}^{2}}{4\pi}\sum_{\alpha\beta}\mathrm{Tr}\left(
\frac{\partial\hat{S}_{\alpha\beta}^{\dagger}}{\partial r_{w}}\frac
{\partial\hat{S}_{\alpha\beta}}{\partial r_{w}}\right)  \,. \label{CP-IE}%
\end{equation}
Evaluating the scattering-matrix expressions on the right-hand side of the
above equations leads to microscopic magnetotransport response coefficients
that describe the interaction of the domain wall with electric currents,
including spin transfer and pumping effects.

These results leads to microscopic expressions for the phenomenological
response\cite{Hals:prl09} of the domain-wall velocity $\dot{r}_{w}$ and charge
current $I^{c}$ to a voltage $V$ and magnetic field applied along the wire
$H$:
\begin{equation}
\left(
\begin{array}
[c]{c}%
\dot{r}_{w}\\
I^{c}%
\end{array}
\right)  =\left(
\begin{array}
[c]{cc}%
L_{ww} & L_{wc}\\
L_{cw} & L_{cc}%
\end{array}
\right)  \left(
\begin{array}
[c]{c}%
2AM_{s}H\\
V
\end{array}
\right)  \,, \label{CP-LL}%
\end{equation}
subject to appropriate conventions for the signs of voltage and magnetic field
and assuming a head-to-head or tail-to-tail wall such that the magnetization
outside of the wall region is collinear with the wire axis. $2AM_{s}H\ $is the
thermodynamic force normalized to the entropy production by the magnetic
system, where $A$ is the cross-sectional area of the wire. We may therefore
expect the Onsager's symmetry relation $L_{cw}=L_{wc}$. When a magnetic field
moves the domain wall in the absence of a voltage $I^{c}=(L_{cw}/L_{ww}%
)\dot{r}_{w}$, which, according to Eq.~(\ref{CP-Ic}) leads to the ratio
$L_{cw}/L_{ww}$ in terms of the scattering matrices. The total energy
dissipation for the same process is $I^{E}=\dot{r}_{w}^{2}/L_{ww}$, which,
according to Eq.~(\ref{CP-IE}), establishes a scattering-matrix expression for
$L_{ww}$ alone. By supplementing these equations with the standard
Landauer-B{\"{u}}ttiker formula for the conductance
\begin{equation}
G=\frac{e^{2}}{h}\mathrm{Tr}\left(  \hat{S}_{12}^{\dagger}\hat{S}_{12}\right)
\,,
\end{equation}
valid in the absence of domain-wall dynamics, we find $L_{cc}$ in the same
spirit since $G=L_{cc}-L_{wc}^{2}/L_{ww}$. Summarizing, the phenomenological
response coefficients in Eq.~(\ref{CP-LL}) read\cite{Hals:prl09}:
\begin{align}
L_{ww}^{-1}  &  =\frac{\hbar}{4\pi}\sum_{\alpha\beta}\mathrm{Tr}\left(
\frac{\partial\hat{S}_{\alpha\beta}^{\dagger}}{\partial r_{w}}\frac
{\partial\hat{S}_{\alpha\beta}}{\partial r_{w}}\right)  \,,\\
L_{cw}  &  =L_{wc}=L_{ww}\frac{e}{2\pi}\mathrm{Im}\sum_{\beta}\mathrm{Tr}%
\left(  \frac{\partial\hat{S}_{\alpha\beta}}{\partial r_{w}}\hat{S}%
_{\alpha\beta}^{\dagger}\right)  \,,\\
L_{cc}  &  =\frac{e^{2}}{h}\mathrm{Tr}\left(  \hat{S}_{12}\hat{S}%
_{12}^{\dagger}\right)  +\frac{L_{wc}^{2}}{L_{ww}}\,.
\end{align}

When the wall is sufficiently smooth, we can model spin torques and pumping by
the continuum theory based on the gradient expansion in the magnetic texture,
Eqs.~(\ref{t_stt}) and (\ref{currentpumpingcont}). Solving for the
magnetic-field and current-driven dynamics of such domain walls is then
possible using the Walker ansatz\cite{schryerJAP74,liPRB04dw}. Introducing the
domain-wall width $\lambda_{w}$:
\begin{equation}
\alpha=\frac{\gamma\lambda_{w}}{2AM_{s}L_{ww}}~~~\mathrm{and}~~~\beta
=-\frac{e\lambda_{w}}{\hbar PG}\frac{L_{wc}}{L_{ww}}\,. \label{CP-ab}%
\end{equation}
When the wall is sharp the adiabatic approximation underlying the
leading-order gradient expansion breaks down. These relations can still be
used as definitions of the effective domain-wall $\alpha$ and $\beta$. As
such, these could be distinct from the bulk values that are associated with
smooth textures. This is relevant for dilute magnetic semiconductors, for which the
adiabatic approximation easily breaks down\cite{Hals:prl09}. In
transition-metal ferromagnets, on the other hand, the adiabatic approximation
is generally perceived to be a good starting point, and we may expect the
dissipative parameters in Eq.~(\ref{CP-ab}) to be comparable to their bulk
values discussed in Sec.~\ref{C-STT}.

\section{First-principles Calculations}

\label{abinitio}

We have shown that the essence of spin pumping and spin transfer can be
captured by a small number of phenomenological parameters. In this section we
address the material dependence of these phenomena in terms of the
(reflection) mixing conductance $G_{\perp}$, the dimensionless Gilbert damping
parameter $\alpha$, and the out-of-plane torque parameter $\beta$.

For discrete systems the (reflection) mixing conductance $G_{\perp}$ was
studied theoretically by Xia \textit{et al}.\cite{Xia:prb02}, Zwierzycki
\textit{et al}.\cite{Zwierzycki:prb05} and Carva \textit{et al}%
.\cite{Carva:prb07}. $G_{\perp}$ describes the spin current flowing in
response to an externally applied spin accumulation $e\mathbf{V}_s$ that is a
vector with length equal to half of the spin-splitting of the chemical
potentials $e|\mathbf{V}_s|=e(V_{\uparrow}-V_{\downarrow})/2$. It also
describes the spin torque exerted on the moment of the magnetic layer
\cite{Slonczewski:jmmm96,Brataas:prl00,Stiles:prb02,Xia:prb02,Brataas:prl03,Zwierzycki:prb05,Carva:prb07}%
. Consider a spin accumulation in a normal metal $N$, which is in contact with
a ferromagnet on the right magnetized along the $z$ axis. The spin current
incident on the interface is proportional to the number of incident channels
in the left lead, $\mathbf{I}_{\mathrm{in}}^{\mathrm{N}}=2G_{\mathrm{N}}^{\mathrm{Sh}} \mathbf{V}_s$, while the reflected spin current is
given by
\begin{equation}
\mathbf{I}_{\mathrm{out}}^{\mathrm{N}}=2\left(
\begin{array}
[c]{ccc}%
G_{\mathrm{N}}^{\mathrm{Sh}}-G_{\perp}^{(R)} & -G_{\perp}^{(I)} & 0\\
G_{\perp}^{(I)} & G_{\mathrm{N}}^{\mathrm{Sh}}-G_{\perp}^{(R)} & 0\\
0 & 0 & G_{\mathrm{N}}^{\mathrm{Sh}}-\frac{G_{\uparrow}+G_{\downarrow}}{2}%
\end{array}
\right)  \mathbf{V}_s \, ,
\label{sc_reflect}%
\end{equation}
where $G_{\sigma}$ are the conventional Landauer-B\"{u}ttiker conductances.
The real and imaginary parts of $G_{\mathrm{N}}^{\mathrm{Sh}}-G_{\perp}%
=(e^2/h) \sum_{mn}r_{mn}^{\uparrow}r_{mn}^{\downarrow\star}$ are related to the
components of the reflected transverse spin current and can be calculated by
considering a single N$|$F interface\cite{Xia:prb02}. When the ferromagnet is
a layer with finite thickness $d$ sandwiched between normal metals, the
reflection mixing conductance depends on $d$ and it is necessary to consider
also the transmission mixing conductance $(e^2/h)\sum_{mn}t_{mn}^{\prime\uparrow
}t_{mn}^{\prime\downarrow\star}$. In Ref.~\onlinecite{Zwierzycki:prb05}, both
reflection and transmission mixing conductances were calculated for Cu$|$%
Co$|$Cu and Au$|$Fe$|$Au sandwiches as a function of magnetic layer thickness
$d$. The real and imaginary parts of the transmission mixing conductance and
the imaginary part of the reflection mixing conductance were shown to decay
rapidly with increasing $d$ implying that the absorption of the transverse
component of the spin current occurs within a few monolayers of the N$|$F
interface for ideal lattice matched interfaces. When a minimal amount of
interface disorder was introduced the absorption increased. The limit
$G_{\perp}\rightarrow G_{\mathrm{N}}^{\mathrm{Sh}}$ corresponds to the
situation where all of the incoming transverse polarized spin current is
absorbed in the magnetic layer. The torque is then proportional to the Sharvin
conductance of the normal metal. This turns out to be the situation for all
but the thinnest (few monolayers) and cleanest Co and Fe magnetic layers
considered by Zwierzycki \textit{et al}.\cite{Zwierzycki:prb05} However, when
there is nesting between Fermi surface sheets for majority and minority spins
so that both spins have the same velocities over a large region of reciprocal
space, then the transverse component of the spin current does not damp so
rapidly and $G_{\perp}$ can continue to oscillate for large values of $d$.
This has been found to occur for ferromagnetic Ni in the (001)
direction.\cite{Carva:prb07}


Eq. \ref{torque_pump} implies that the spin pumping renormalizes both the
Gilbert damping parameter $\alpha$ and the gyromagnetic ratio $\gamma$ of a
ferromagnetic film embedded in a conducting non-magnetic medium. However, in
view of the results discussed in the previous paragraph, we conclude that the
main effect of the spin pumping is to enhance the Gilbert damping. The
correction is directly proportional to the real part of the reflection mixing
conductance and is essentially an interface property. Oscillatory effects are
averaged out for realistic band structures, especially in the presence of
disorder. $G_{\perp}^{(R)}$ determines the damping enhancement of a single
ferromagnetic film embedded in a perfect spin-sink medium and is usually very
close to $G_{\mathrm{N}}^{\mathrm{Sh}}$ for intermetallic
interfaces\cite{Xia:prb02,Stiles:prb02}.

\subsection{Alpha}

We begin with a discussion of the small-angle damping measured as a function
of temperature using ferromagnetic resonance (FMR). There is general agreement
that spin-orbit coupling and disorder are essential ingredients in any
description of how spin excitations relax to the ground state. In the absence
of intrinsic disorder, one might expect the damping to increase monotonically
with temperature in clean magnetic materials and indeed, this is what is
observed for Fe. Heinrich \textit{et al.} \cite{Heinrich:pssb67} developed an
explicit model for this high-temperature behaviour in which itinerant $s$
electrons scatter from localized $d$ moments and transfer spin angular
momentum to the lattice via spin-orbit interaction. This $s-d$ model results
in a damping that is inversely proportional to the electronic relaxation time,
$\alpha\sim1/\tau$, i.e., is \textit{resistivity}-like. However, at low
temperatures, both Co and Ni exhibit a sharp rise in damping as the
temperature decreases. The so-called breathing Fermi surface model was
proposed \cite{Kambersky:cjp70,Korenman:prb72,Kunes:prb02} to describe this
low-temperature \textit{conductivity}-like damping, $\alpha\sim\tau$. In this
model the electronic population lags behind the instantaneous equilibrium
distribution due to the precessing magnetization and requires dissipation of
energy and angular momentum to bring the system back to equilibrium.

Of the numerous microscopic models that have been proposed \cite{Heinrich:05}
to explain the damping behaviour of metals, only the so-called ``torque
correlation model'' (TCM) \cite{Kambersky:cjp76} is qualitatively successful
in explaining the non-monotonic damping observed for hcp Co that results from
conductivity-like and resistivity-like behaviours at low and high
temperatures, respectively. The central result of the TCM is the expression
\begin{equation}
\tilde G = \frac{g^{2} \mu_{B}^{2}}{\hbar}\sum_{n,m} \int\frac{d \mathbf{k}%
}{(2\pi)^{3}} \left|  {\langle{n,\mathbf{k}} \vert[ \sigma_{-},\hat{\cal H}_{so}] \vert{m,\mathbf{k}} \rangle} \right|  ^{2}
W_{n,m}(\mathbf{k}) \label{eqn:tcm}%
\end{equation}
for the damping. The commutator $[\sigma_{-},\hat{\cal H}_{so}]$ describes a torque between the spin and orbital moments that
arises as the spins precess. The corresponding matrix elements in
\eqref{eqn:tcm} describe transitions between states in bands $n$ and $m$
induced by this torque whereby the crystal momentum $\mathbf{k}$ is conserved.
Disorder enters in the form of a phenomenological relaxation time $\tau$ via
the spectral overlap
\begin{equation}
W_{n,m}(\mathbf{k}) = -\frac{1}{\pi} \int A_{n}(\varepsilon,\mathbf{k})
A_{m}(\varepsilon,\mathbf{k}) \frac{df}{d\varepsilon} d\varepsilon
\label{eqn:spo}%
\end{equation}
where the electron spectral function $A_{n}(\varepsilon,\mathbf{k}) $ is a
Lorentzian centred on the band $n$, whose width is determined by the
scattering rate. For intraband transitions with $m=n$, integration over energy
yields a spectral overlap which is proportional to the relaxation time, like
the conductivity. For interband transitions with $m \ne n$, the energy
integration leads to a spectral overlap that is roughly inversely proportional
to the relaxation time, like the resistivity.

To interpret results obtained with the TCM, Gilmore et
al.\cite{Gilmore:prl07,Gilmore:jap08,Gilmore:prb10,Kambersky:cjp76,Kambersky:prb07}
used an effective field approach expressing the effective field about which
the magnetization precesses in terms of the total energy
\begin{equation}
\mu_{0} \mathbf{H}^{\mathrm{eff}} = - \frac{\partial E}{\partial\mathbf{M}}%
\end{equation}
and then approximated the total energy by a sum of single particle eigenvalues
$E \sim\sum_{n,\mathbf{k}} \varepsilon_{n \mathbf{k}} f_{n \mathbf{k}}$, so
that the effective field naturally splits into two parts
\begin{equation}
\mathbf{H}^{\mathrm{eff}} = \frac{1}{\mu_{0} M} \sum_{n,\mathbf{k}} \left[
\frac{\partial\varepsilon_{n \mathbf{k}}}{\partial\mathbf{m}} f_{n \mathbf{k}}
+ \varepsilon_{n \mathbf{k}} \frac{\partial f_{n \mathbf{k}}}{\partial
\mathbf{m}} \right]  
\label{eqn:efff}%
\end{equation}
the first of which corresponds to the breathing Fermi surface model, intraband
transitions and conductivity-like behaviour while the second term could be
related to interband transitions and resistivity-like behaviour. Evaluation of
this model for Fe, Co and Ni using first-principles calculations to determine
$\varepsilon_{n \mathbf{k}} $ including spin-orbit coupling yields results for
the damping $\alpha$ in good qualitative and reasonable quantitative agreement
with the experimental observations.\cite{Gilmore:prl07}

In spite of this real progress, the TCM has disadvantages. As currently
formulated, the model can only be applied to periodic lattices. Extending it
to handle inhomogeneous systems such as ferromagnetic substitutional alloys
like Permalloy (Ni$_{80}$Fe$_{20}$), magnetic multilayers or heterojunctions,
disordered materials or materials with surfaces is far from trivial. The TCM
incorporates disorder in terms of a relaxation time parameter $\tau$ and so
suffers from the same disadvantages as all transport theories similarly
formulated, namely, that it is difficult to relate microscopically measured
disorder unambiguously to a given value of $\tau$. Indeed, since $\tau$ in
general depends on incoming and scattered band index $n$, wave vector
$\mathbf{k}$, as well as spin index, assuming a single value for it is a gross
simplification. A useful theoretical framework should allow us to study not
only crystalline materials such as the ferromagnetic metals Fe, Co and Ni and
substitutional disordered alloys such as permalloy (Py), but also amorphous
materials and configurations such as magnetic heterojunctions, multilayers,
thin films etc. which become more important and are more commonly encountered
as devices are made smaller.

The scattering theoretical framework discussed in section IIIB satisfies these
requirements and has recently been implemented by extending a first-principles
scattering formalism \cite{Xia:prb01,Xia:prb06} based upon the local spin
density approximation (LSDA) of density functional theory (DFT) to include
non-collinearity, spin-orbit coupling (SOC) and chemical or thermal disorder
on equal footings.\cite{Starikov:prl10} Relativistic effects are included by
using the Pauli Hamiltonian. To calculate the scattering matrix, a
``wave-function matching'' (WFM) scheme \cite{Ando:prb91,Xia:prb01,Xia:prb06}
implemented with a minimal basis of tight-binding linearized muffin-tin
orbitals (TB-LMTOs) \cite{Andersen:prb86,Andersen:prb75}.
Atomic-sphere-approximation (ASA) potentials
\cite{Andersen:prb86,Andersen:prb75} are calculated self-consistently using a
surface Green's function (SGF) method also implemented \cite{Turek:97} with TB-LMTOs.

\subsubsection{NiFe alloys.}

The flexibility of the scattering theoretical formulation of transport can be
demonstrated with an application to NiFe binary alloys.\cite{Starikov:prl10}
Charge and spin densities for binary alloy $A$ and $B$ sites are calculated
using the coherent potential approximation (CPA) \cite{Soven:pr67} generalized
to layer structures \cite{Turek:97}. For the transmission matrix calculation,
the resulting spherical potentials are distributed at random in large lateral
supercells (SC) subject to maintenance of the appropriate concentration of the
alloy \cite{Xia:prb01,Xia:prb06}. Solving the transport problem using lateral
supercells makes it possible to go beyond effective medium approximations such
as the CPA. As long as one is only interested in the properties of bulk
alloys, the leads can be chosen for convenience and Cu leads with a single
scattering state for each value of crystal momentum, $\mathbf{k}_{\parallel}$
are very convenient. The alloy lattice constants are determined using Vegard's
law and the lattice constants of the leads are made to match. Though NiFe is
fcc only for the concentration range $0 \leq x \leq0.6$, the fcc structure is
used for all values of $x$.

\begin{figure}[ptb]
\includegraphics[width=0.5\columnwidth]{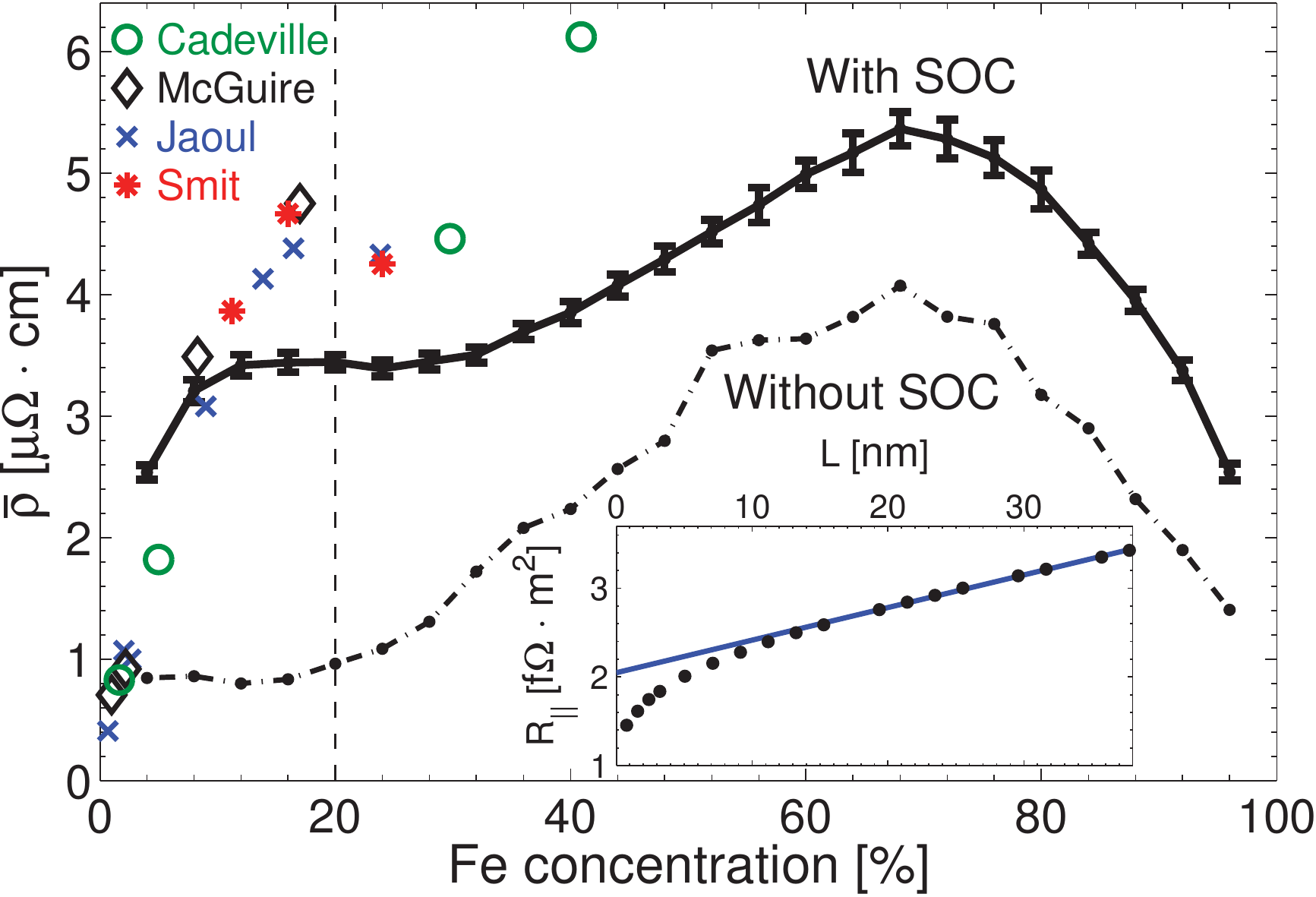}\caption{ Calculated
resistivity as a function of the concentration $x$ for fcc Ni$_{1-x}$Fe$_{x}$
binary alloys with (solid line) and without (dashed-dotted line) SOC. Low
temperature experimental results are shown as symbols
\cite{Smit:phys51,McGuire:ieeem75,Jaoul:jmmm77,Cadeville:jpf73}. The
composition Ni$_{80}$Fe$_{20}$ is indicated by a vertical dashed line. Inset:
resistance of Cu$|$Ni$_{80}$Fe$_{20}|$Cu as a function of the thickness of the
alloy layer. Dots indicate the calculated values averaged over five
configurations while the solid line is a linear fit. }%
\label{fig:Fig1}%
\end{figure}

To illustrate the methodology, we begin by calculating the electrical
resistivity of Ni$_{80}$Fe$_{20}$. In the Landauer-B{\"ut}tiker formalism, the
conductance can be expressed in terms of the transmission matrix $t$ as $G =
(e^{2}/h) Tr \left\{  tt^{\dag}\right\}  $ \cite{Buttiker:prb85,Datta:95}. The
resistance of the complete system consisting of ideal leads sandwiching a
layer of ferromagnetic alloy of thickness $L$ is $R(L) = 1/G(L) =
1/G_{\mathrm{Sh}} + 2 R_{\mathrm{if}} + R_{\mathrm{b}}(L) $ where
$G_{\mathrm{Sh}}= \left(  2 e^{2}/h \right)  N $ is the Sharvin conductance of
each lead with $N$ conductance channels per spin, $R_{\mathrm{if}}$ is the
interface resistance of a single N$|$F interface, and $R_{\mathrm{b}}(L)$ is
the bulk resistance of a ferromagnetic layer of thickness $L$
\cite{Schep:prb97,Xia:prb06}. When the ferromagnetic slab is sufficiently
thick, Ohmic behaviour is recovered whereby $R_{\mathrm{b}}(L)\approx\rho L$
as shown in the inset to Fig.~\ref{fig:Fig1} and the bulk resistivity $\rho$
can be extracted from the slope of $R(L)$. For currents parallel and
perpendicular to the magnetization direction, the resistivities are different
and have to be calculated separately. The average resistivity is given by
$\bar{\rho}=(\rho_{\parallel}+2\rho_{\perp})/3$, and the anisotropic
magnetoresistance ratio (AMR) by $(\rho_{\parallel}-\rho_{\perp})/\bar{\rho}$.

For Ni$_{80}$Fe$_{20}$ we find values of $\bar{\rho}= 3.5 \pm0.15$~$\mu$Ohm-cm
and AMR $=19 \pm1 \%$, compared to experimental low-temperature values in the
range $4.2-4.8$~$\mu$Ohm-cm for $\bar{\rho}$ and $18\%$ for AMR
\cite{Smit:phys51}. The resistivity calculated as a function of $x$ is
compared to low temperature literature values
\cite{Smit:phys51,McGuire:ieeem75,Jaoul:jmmm77,Cadeville:jpf73} in
Fig.~\ref{fig:Fig1}. The overall agreement with previous calculations is good
\cite{Banhart:epl95,Banhart:prb97}. In spite of the smallness of the SOC, the
resistivity of Py is underestimated by more than a factor of four when it is
omitted, underlining its importance for understanding transport properties.

\begin{figure}[ptb]
\includegraphics[width=0.5\columnwidth]{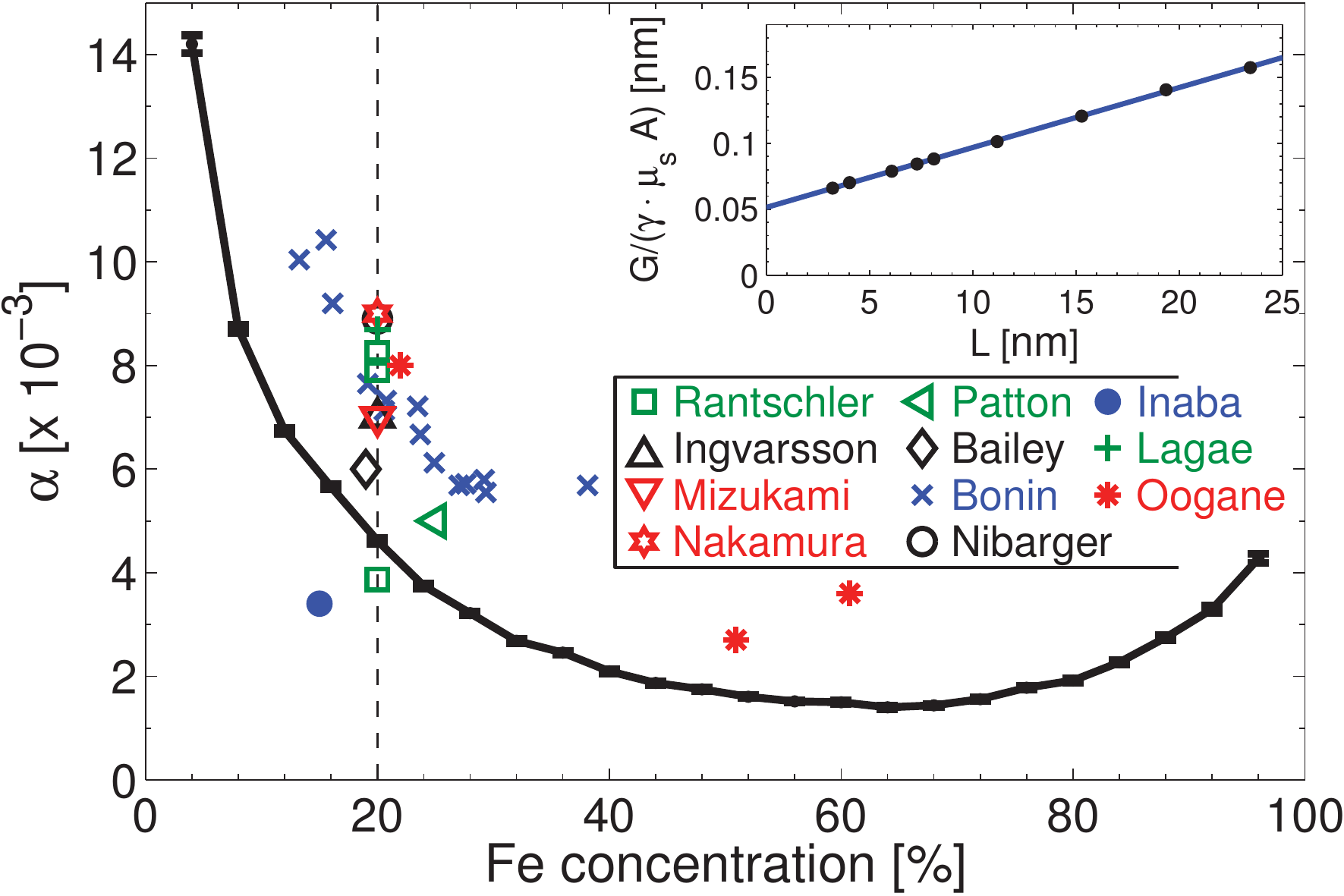}\caption{Calculated
zero temperature (solid line) and experimental room temperature (symbols)
values of the Gilbert damping parameter as a function of the concentration $x$
for fcc Ni$_{1-x}$Fe$_{x}$ binary alloys
\cite{Mizukami:jmmm01,Mizukami:jjap01,Urban:prl01,Heinrich:prl03,Bailey:ieeem01,
Patton:jap75,Ingvarsson:apl04,Nakamura:jjap04,Rantschler:ieeem05,Bonin:jap05,
Lagae:jmmm05,Nibarger:apl03,Inaba:ieeem06,Oogane:jjap06}. Inset: total damping
of Cu$|$Ni$_{80}$Fe$_{20}|$Cu as a function of the thickness of the alloy
layer. Dots indicate the calculated values averaged over five configurations
while the solid line is a linear fit. }%
\label{fig:Fig2}%
\end{figure}

Assuming that the Gilbert damping is isotropic for cubic substitutional alloys
and allowing for the enhancement of the damping due to the F$|$N interfaces
\cite{Tserkovnyak:prl02,Zwierzycki:prb05,Mizukami:jmmm01,Mizukami:jjap01,Urban:prl01,Heinrich:prl03}, the
total damping in the system with a ferromagnetic slab of thickness $L$ can be
written ${\tilde G}(L)={\tilde G}_{\mathrm{if}}+{\tilde G}_{b}(L) $ where we
express the bulk damping in terms of the dimensionless Gilbert damping
parameter ${\tilde G_{b}}(L)=\alpha\gamma M_{s}(L) = \alpha\gamma\mu_{s} A L$,
where $\mu_{s}$ is the magnetization density and $A$ is the cross section. The
results of calculations for Ni$_{80}$Fe$_{20}$ are shown in the inset to
Fig.~\ref{fig:Fig2}. The intercept at $L=0$, ${\tilde G}_{\mathrm{if}}$,
allows us to extract the damping enhancement \cite{Zwierzycki:prb05} but here
we focus on the bulk properties and leave consideration of the material
dependence of the interface enhancement for later study. The value of $\alpha$
determined from the slope of ${\tilde G}(L)/(\gamma\mu_{s} A)$ is $0.0046
\pm0.0001$ that is at the lower end of the range of values $0.004 - 0.013$
measured at room temperature for Py
\cite{Mizukami:jmmm01,Mizukami:jjap01,Urban:prl01,Heinrich:prl03,Bailey:ieeem01,
Patton:jap75,Ingvarsson:apl04,Nakamura:jjap04,Rantschler:ieeem05,Bonin:jap05,
Lagae:jmmm05,Nibarger:apl03,Inaba:ieeem06,Oogane:jjap06}.

Fig.~\ref{fig:Fig2} shows the Gilbert damping parameter as a function of $x$
for Ni$_{1-x}$Fe$_{x}$ binary alloys in the fcc structure. From a large value
for clean Ni, it decreases rapidly to a minimum at $x \sim0.65$ and then grows
again as the limit of clean \emph{fcc} Fe is approached. Part of the decrease
in $\alpha$ with increasing $x$ can be explained by the increase in the
magnetic moment per atom as we progress from Ni to Fe. The large values of
$\alpha$ calculated in the dilute alloy limits can be understood in terms of
conductivity-like enhancement at low temperatures
\cite{Bhagat:prb74,Heinrich:jap79} that has been explained in terms of
intraband scattering
\cite{Kambersky:cjp76,Gilmore:prl07,Gilmore:jap08,Kambersky:prb07}. The trend
exhibited by the theoretical $\alpha(x)$ is seen to be reflected by
experimental results obtained at room temperature. In spite of a large spread
in measured values, these seem to be systematically larger than the calculated
values. Part of this discrepancy can be attributed to an increase in $\alpha$
with temperature \cite{Bastian:pssa76,Bailey:ieeem01}.


Calculating $\alpha$ for the end members, Ni and Fe, of the substitutional
alloy Ni$_{1-x}$Fe$_{x}$ presents a practical problem. In these limits there
is no scattering whereas in experiment there will always be some residual
disorder at low temperatures and at finite temperatures, electrons will
scatter from the thermally displaced ions. We introduce a simple ``frozen
thermal disorder'' scheme to study Ni and Fe and simulate the effect of
temperature via electron-phonon coupling by using a random Gaussian
distribution of ionic displacements $\mathbf{u}_{i}$, corresponding to a
harmonic approximation. This is characterized by the root-mean-square (RMS)
displacement $\Delta=\sqrt{\langle\vert{\mathbf{u}}_{i}\vert^{2}\rangle}$
where the index $i$ runs over all atoms. Typical values will be of the order
of a few hundredths of an angstrom. We will not attempt to relate $\Delta$ to
a real lattice temperature here.

\begin{figure}[ptb]
\includegraphics[width=0.5\columnwidth]{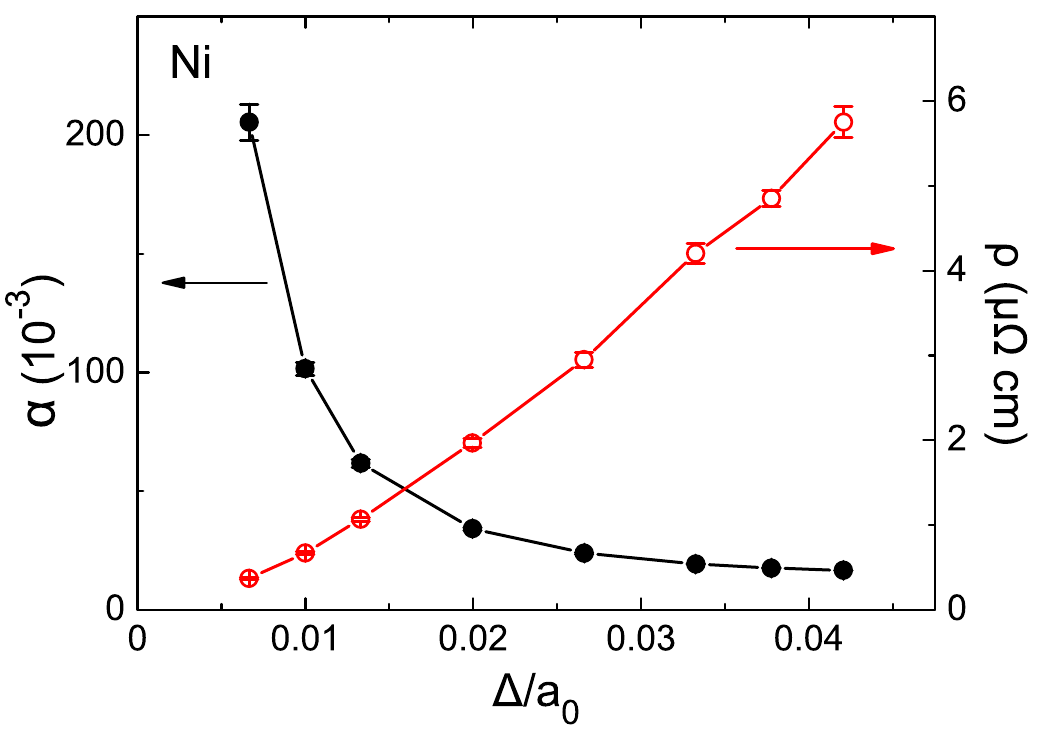}\caption{Calculated
Gilbert damping and resistivity for fcc Ni as function of the relative RMS
displacement with respect to the corresponding lattice constant, $a_{0}
=3.524$ \AA .}%
\label{fig3}%
\end{figure}

We calculate the total resistance $R(L)$ and Gilbert damping ${\tilde G}(L)$
for thermally disordered scattering regions of variable length $L$ and extract
the resistivity $\rho$ and damping $\alpha$ from the slopes as before. The
results for Ni are shown as a function of the RMS displacement in
Fig.~\ref{fig3}. The resistivity is seen to increase monotonically with
$\Delta$ underlining the correlation between $\Delta$ and a real temperature.
For large values of $\Delta$, $\alpha$ saturates for Ni in agreement with
experiment \cite{Bhagat:prb74} and calculations based on the
torque-correlation model \cite{Kambersky:prb07,Gilmore:prl07,Gilmore:prb10}
where no concrete scattering mechanism is attached to the relaxation time
$\tau$. The absolute value of the saturated $\alpha$ is about 70\% of the
observed value. For small values of $\Delta$, the Gilbert damping increases
rapidly as $\Delta$ decreases. This sharp rise corresponds to the
experimentally observed conductivity-like behaviour at low temperatures and
confirms that the scattering formalism can reproduce this feature.

\subsection{Beta}

\begin{figure}[t]
\begin{center}
\includegraphics[width=0.5\columnwidth]{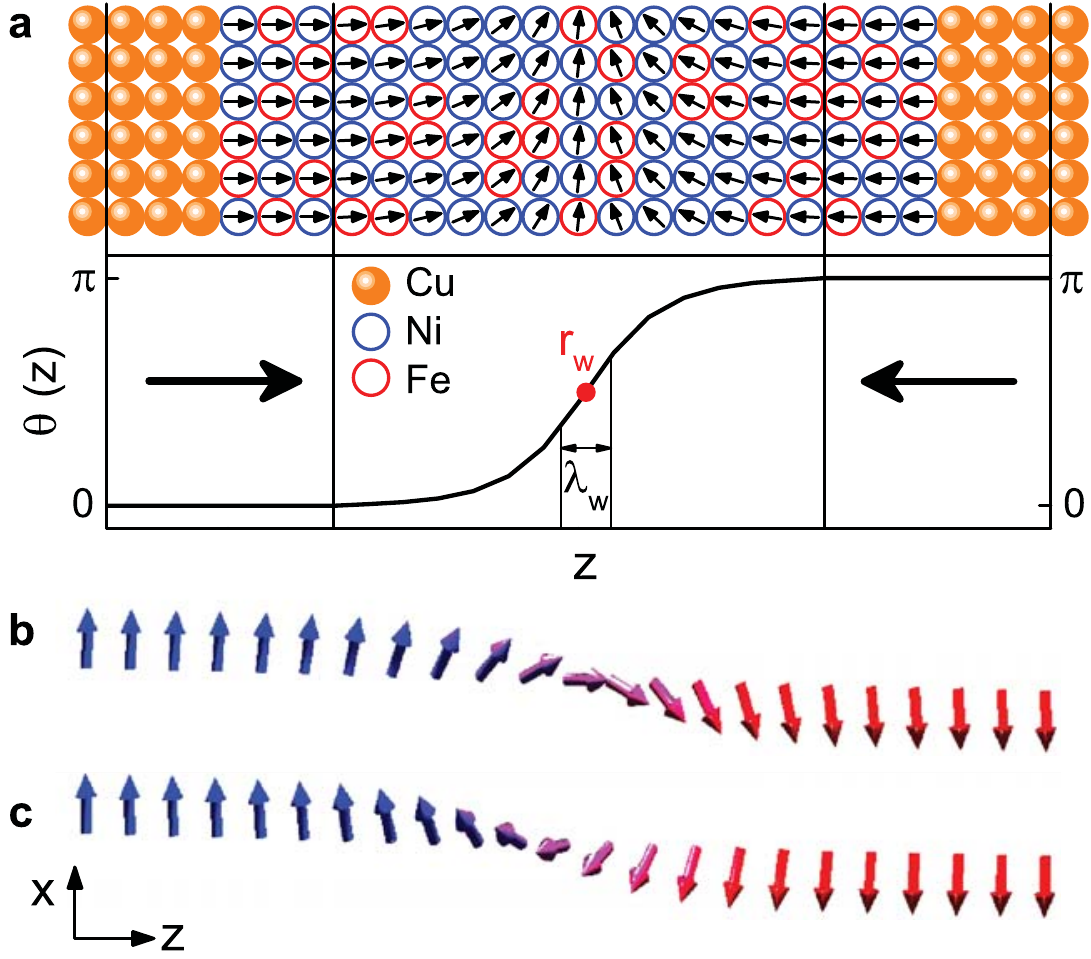}
\end{center}
\caption{(a) Sketch of the configuration of a N{\'e}el DW in Py sandwiched by
two Cu leads. The arrows denotes local magnetization directions. The curve
shows the mutual angle between the local magnetization and the transport
direction ($z$ axis). (b) Magnetization profile of the rotated N{\'e}el wall.
(c) Magnetization profile of the Bloch wall.}%
\label{fig:model}%
\end{figure}

To evaluate expressions \eqref{CP-ab} for the out-of-plane
spin-torque parameter $\beta$ given in Section IIIB requires modelling domain
walls (DW) in the scattering region sandwiched between ideal Cu leads. A
head-to-head N{\'e}el DW is introduced inside the permalloy region by rotating
the local magnetization to follow the Walker profile, $\mathbf{m}%
(z)=[f(z),0,g(z)]$ with $f(z)= \cosh^{-1}[(z-r_{w})/\lambda_{w}]$ and
$g(z)=-\tanh[(z-r_{w})/\lambda_{w}]$ as shown schematically in
Fig.~\ref{fig:model}(a). $r_{w}$ is the DW center and $\lambda_{w}$ is a
parameter characterizing its width. In addition to the N{\'e}el wall, we also
study a rotated N{\'e}el wall with magnetization profile $\mathbf{m}%
(z)=[g(z),0,f(z)]$ sketched in Fig.~\ref{fig:model}(b) and a Bloch wall with
$\mathbf{m}(z)=[g(z),f(z),0]$ sketched in Fig.~\ref{fig:model}(c).

\begin{figure}[b]
\begin{center}
\includegraphics[width=0.5\columnwidth]{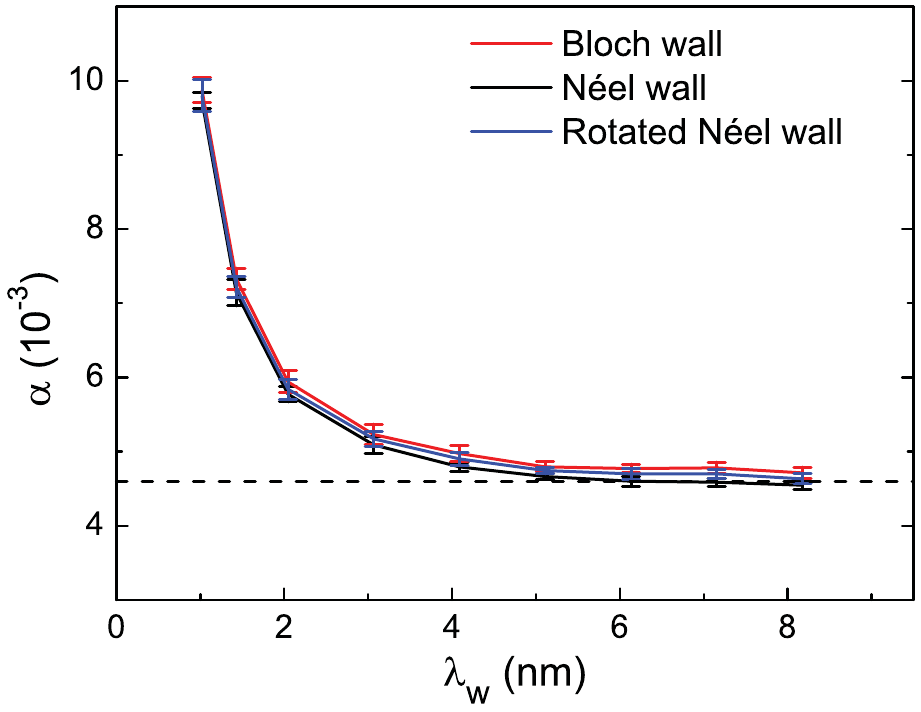}
\end{center}
\caption{Calculated effective Gilbert damping constant $\alpha$ for Py DWs as
a function of $\lambda_{w}$. The dashed lines show the calculated $\alpha$ for
bulk Py with the magnetization parallel to the transport direction
\cite{Starikov:prl10}. }%
\label{fig:alpha}%
\end{figure}

The effective Gilbert damping constant $\alpha$ of permalloy in the presence
of all three DWs calculated using \eqref{CP-ab} is shown in
Fig.~\ref{fig:alpha}. For different types of DWs, $\alpha$ is identical within
the numerical accuracy indicating that the Gilbert damping is isotropic due to
the strong impurity scattering \cite{Gilmore:prb10}. In the adiabatic limit,
$\alpha$ saturates to the same value (the dashed lines in Fig.~\ref{fig:alpha}%
) calculated for bulk permalloy using \eqref{Gilbertscattering}. It implies that the
DWs in permalloy have little effect on the magnetization relaxation and the
strong impurity scattering is the dominant mechanism to release energy and
magnetization. This is in contrast to DWs in (Ga,Mn)As where Gilbert damping
is mostly contributed by the reflection of the carriers from the DW.
\cite{Hals:prl09} At $\lambda_{w}<5$~nm, the non-adiabatic reflection of
conduction electrons due to the rapidly-varying magnetization direction
becomes significant and results in a sharp rise in $\alpha$ for narrow DWs.

\begin{figure}[ptb]
\begin{center}
\includegraphics[width=0.5\columnwidth]{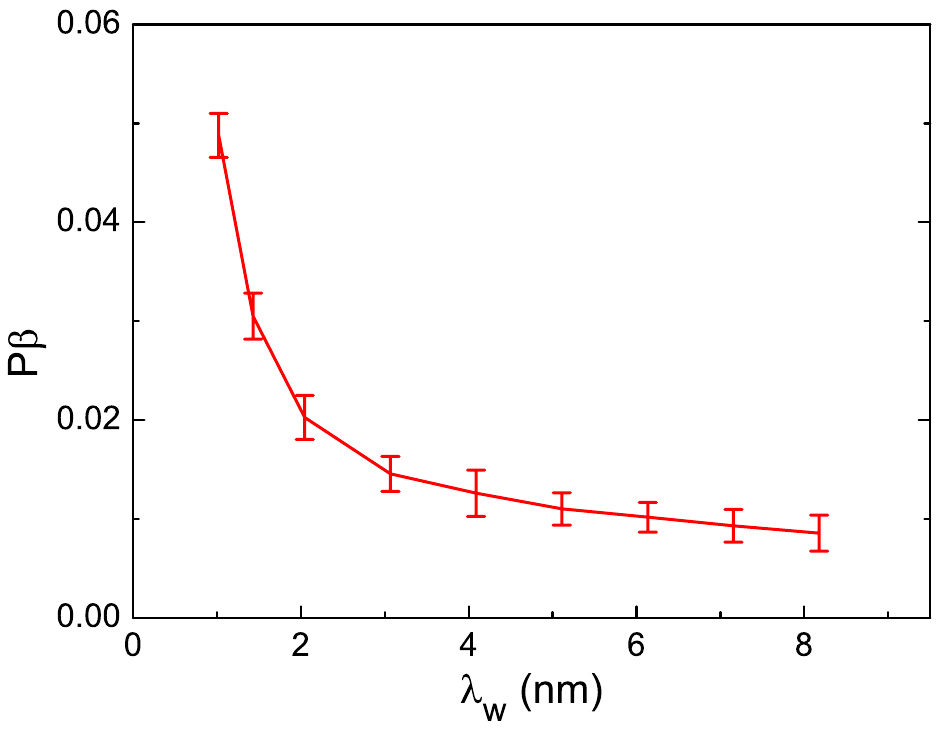}
\end{center}
\caption{Calculated out-of-plane spin torque parameter $P\beta$ for permalloy
DWs as a function of $\lambda_{w}$. }%
\label{fig:beta}%
\end{figure}

The out-of-plane torque is formulated as $\beta(\hbar\gamma P/2eM_{s}%
)\mathbf{m}\times(\mathbf{j}\cdot\nabla)\mathbf{m}$ in the
Landau-Lifshitz-Gilbert (LLG) equation under a finite current density
$\mathbf{j}$. In principle, the current polarization $P$ is required to
determine $\beta$. Since the spin-dependent conductivities of permalloy depend
on the angle between the current and the magnetization, $P$ is not
well-defined for magnetic textures. Instead, we calculate the quantity
$P\beta$, as shown in Fig.~\ref{fig:beta} for a Bloch DW. For $\lambda_{w}%
<5$~nm, $P\beta$ decreases quite strongly with increasing $\lambda_{w}$
corresponding to an expected non-adiabatic contribution to the out-of-plane
torque. This arises from the spin-flip scattering induced by the
rapidly-varying magnetization in narrow DWs\cite{Xiao:prb06} and does not
depend on the specific type of DW. For $\lambda_{w}>5$~nm, which one expects
to be in the adiabatic limit, $P\beta$ decreases slowly to a constant value
\cite{Tatara:prp08,Hayashi:apl08,Lepadatu:prl09,Lepadatu:prb10,Moore:prb09,Eltschka:prl10,Adam:prb09,Burrowes:natp10,Zhang:prl04,Thiaville:epl05,Xiao:prb06,Kohno:jopj06,Garate:prb09,Hals:prl09}%
. It is unclear what length scale is varying so slowly. Unfortunately, the
spread of values for different configurations is quite large for the last data
point and our best estimate of $P\beta$ for a Bloch DW in permalloy is
$\sim0.08$. Taking the theoretical value of $P \sim0.7$ for permalloy
\cite{Starikov:prl10}, our best estimate of $\beta$ is a value of $\sim0.01$.

\section{Theory versus Experiments}

Spin-torque induced magnetization dynamics in multilayers and its reciprocal
effect, the spin pumping, are experimentally well established and
quantitatively understood within the framework described in this paper, and
need not to be discussed further here.\cite{Tserkovnyak:rmp05,Ralph:jmmm08}
Recent FMR experiments also confirm the spin-pumping contribution to the enhanced magnetization dissipation\cite{Ghosh:apl11}. Spin-pumping occurs in magnetic insulators as well\cite{Kajiwara,Sandweg:prl11}.

The parameters that control the current-induced dynamics of continuous
textures are much less well known. Most experiments are carried out on
permalloy (Py). It is a magnetically very soft material with large domain wall
widths of the order of 100 nm. Although the adiabatic approximations appears
to be a safe assumption in Py, many systems involve vortex domain walls with
large gradients in the wall center, and, therefore, possibly sizable
nonadiabatic corrections. Effective description for such vortex dynamics has
been constructed in Ref.~\onlinecite{Wong:prb10}, where it was shown, in
particular, that self-consistent quadratic corrections to damping (which stem
from self-pumped currents inducing backaction on the magnetic order) is
generally non-negligible in transition-metal ferromagnets.

Early experimental studies\cite{Hayashi06,Meier07} for the torque-supplemented
[Eq.~(\ref{t_stt})] LLG equation describing current-driven domain-wall motion
in magnetic wires reported values of the $\beta/\alpha$ ratios in Py close to
unity, in agreement with simple Stoner-model calculations. However, much
larger values $\beta/\alpha\sim8$ was extracted from the current-induced
oscillatory motion of domain walls.\cite{thomasNAT06} The inequality$\beta
\neq\alpha$ was also inferred from a characteristic transverse to vortex wall
structure transformation, although no exact value of the ratio was
established.\cite{heynePRL08} In Ref.~\cite{Lepadatu}, vanadium doping of Py
was shown to enhance $\beta$ up to nearly 10$\alpha$, with little effect on
$\alpha$ itself. Even larger ratios, $\beta/\alpha\sim20$, were found for
magnetic vortex motion by an analysis of their displacement as a function of
an applied dc current in disc structures.\cite{Kruger,Heyne10}

Eltschka \textit{et~al.}\cite{eltschka10} reported on a measurement of the
dissipative spin-torque parameter $\beta$ entering Eq.~(\ref{t_stt}), as
manifested by a thermally-activated motion of transverse and vortex domain
walls in Py. They found the ratio $\beta_{v}/\beta_{t}\sim7$ for the vortex vs
transverse wall, attributing the larger $\beta$ to high magnetization
gradients in the vortex wall core. Their ratio $\beta_{t}/\alpha\sim1.3$ turns
out to be close to unity, where $\alpha$ is the bulk Gilbert damping. The
importance of large spin-texture gradients on the domain-wall and vortex
dynamics was theoretically discussed in Refs.~\onlinecite{Foros:prb08,Wong:prb10}.

The material dependence of the current-induced torques is not yet well
investigated. A recent study on CoNi and FePt wires with perpendicular
magnetization found $\beta\approx\alpha$, in spite of the relatively narrow
domain walls in these materials.\cite{Burrowes} Current-induced domain-wall
dynamics in dilute magnetic semiconductors\cite{yamanouchiSCI07} generally
exhibit similar phenomenology, but a detailed discussion, especially of the
domain wall creep regime that can be accessed in these systems, is beyond the
scope of this review.

Finally, the first term in the spin-pumping expression
(\ref{currentpumpingcont}) has been measured by Yang \textit{et~al.}%
\cite{yangPRL09} for a domain wall moved by an applied magnetic field above
the Walker breakdown field. These experiments confirmed the existence of
pumping effects in magnetic textures, which are Onsager reciprocal of spin
torques and thus expected on general grounds. Similar experiments carried out
below the Walker breakdown would also give direct access to the $\beta$ parameter.

\section{Conclusions}

A spin polarized current can excite magnetization dynamics in ferromagnets via
spin-transfer torques. The reciprocal phenomena is spin-pumping where a
dynamic magnetization pumps spins into adjacent conductors. We have discussed
how spin-transfer torques and spin-pumping are directly related by Onsager
reciprocity relations.

In layered normal metal-ferromagnet systems, spin-transfer torques can be
expressed in terms of two conductance parameters governing the flow of spins
transverse to the magnetization direction and the spin-accumulation in the
normal metal. In metallic systems, the field-like torque is typically much
smaller than the effective energy gain/damping torque, but in tunnel systems
they might become comparable. Spin-pumping is controlled by the same
transverse conductance parameters as spin-transfer torques, the magnetization
direction and its rate of change. It can lead to an enhanced magnetization
dissipation in ultra-thin ferromagnets or a build-up of spins, a spin-battery,
in normal metals where the spin-flip relaxation rate is low.

Spin-transfer torque and spin-pumping phenomena in magnetization textures are
similar to their counterparts in layered normal metal-ferromagnet systems. A
current becomes spin polarized in a ferromagnet and this spin-polarized
current in a magnetization texture gives rise to a reactive torque and a
dissipative torque in the lowest gradient expansion. The reciprocal pumping
phenomena can be viewed as an electromotive force, the dynamic magnetization
texture pumps a spin-current that in turn is converted to a charge current or
voltage by the giant magnetoresistance effect. Naturally, the parameters
governing the spin-transfer torques and the pumping phenomena are also the
same in continuously textured ferromagnets.

When the spin-orbit interaction becomes sufficiently strong, additional
effects arise in the coupling between the magnetization and itinerant
electrical currents. A charge potential can then by itself induce a torque on
the ferromagnet and the reciprocal phenomena is that a precessing ferromagnet
can induce a charge current in the adjacent media. The latter can be an
alternative way to carry out FMR measurements on small ferromagnets by
measuring the induced voltage across a normal metal-ferromagnet-normal metal device.

These phenomena are well-know and we have reviewed them in a unified physical
picture and discussed the connection between these and some experimental results.

\acknowledgements

We are grateful to J\o rn Foros, Bertrand I. Halperin, Kjetil M. D. Hals,
Alexey Kovalev, Yi Liu, Hans Joakim Skadsem, Anton Starikov, Zhe Yuan, and
Maciej Zwierzycki for discussions and collaborations.

This work was supported in part by EU-ICT-7 contract no. 257159 MACALO -
Magneto Caloritronics, DARPA, and NSF under Grant No. DMR-0840965.

\end{document}